\documentclass[preprint,12pt]{elsarticle}

\usepackage{amssymb}
\usepackage{amsthm}
\usepackage{amsmath}

\usepackage{longtable}
\usepackage{caption}
\usepackage{subcaption}
\usepackage{threeparttable}
\usepackage[colorlinks=true,linkcolor=blue,citecolor=blue,urlcolor=blue]{hyperref}

\begin{document}
	
\begin{frontmatter}



\title{Kinetic development of low-temperature propane oxidation in a repetitively-pulsed nanosecond discharge}


\author[1]{Zhenyang Li}
\author[2]{Bo Yin}
\author[1]{Qifu Lin}

\author[2]{Yifei Zhu\corref{*}}
\ead{yifei.zhu.plasma@gmail.com}
\cortext[*]{Corresponding author.}

\author[2,3]{Yun Wu}
					
\affiliation[1]{organization={Institute of Energy, Hefei Comprehensive National Science Center},
						city={Hefei},
						postcode={230026},
						country={China}}
\affiliation[2]{organization={National Key Lab of Aerospace Power System and Plasma Technology, Xi'an Jiaotong University},
						city={Xi'an},
						postcode={710049},
						country={China}}
\affiliation[3]{organization={National Key Lab of Aerospace Power System and Plasma Technology, Airforce Engineering University},
						city={Xi'an},
						postcode={710038},
						country={China}}


%

\begin{abstract}
	
	The kinetics of plasma assisted low temperature oxidation of $\rm C_3H_8/O_2/Ar$ mixtures have been studied in a wide specific deposition energy with the help of nanosecond repetitively pulsed discharge. Two types of nanosecond pulsed plasma sources, the nanosecond capillary discharge (nCD) and dielectric barrier discharge (DBD) combined with the synchrotron photoionization mass spectrometry are investigated. The electron impact reaction rate of propane dissociation and some combustion chemical reactions rate constants are updated according to the nCD and DBD experiment results, and uncertainty of the reactions are analyzed in detail. Compared to the existing model, the updated model's prediction accuracy has great improvement in species $\rm H_2O$, $\rm CO$, $\rm CO_2$, $\rm CH_4$, $\rm CH_2O$, $\rm CH_3OH$, $\rm C_2H_2$, $\rm C_2H_4$, $\rm C_2H_6$, $\rm C_2H_5OH$, $\rm C_2H_5OOH$, $\rm C_3H_4$-$\rm A$, $\rm C_3H_4$-$\rm P$, $\rm C_2H_5CHO$, $\rm i$-$\rm C_3H_7OH$ and $\rm C_3H_7OOH$. The propane oxidation processes assisted by DBD and nCD were compared under different single pulse deposition energy (SPDE) conditions while maintaining the same total deposition energy. The reduced electric field in nCD is concentrated at 150-200 Td and 450-500 Td, whereas in DBD it ranges from 0-25 Td and 50-250 Td. Notably, SPDE shows minimal influence on the propane oxidation process, which is primarily controlled by total deposition energy and demonstrates little dependence on the discharge type (DBD or nCD).
		
\end{abstract}



\end{frontmatter}

\section*{Novelty and significance statement}
The kinetics of plasma-assisted low temperature propane oxidation is studied and validated. A kinetic scheme is developed for specific deposited energy per pulse around $5.0 \times 10^{-3}$ eV/mol and reduced electric field ranging from 1 to 500 Td. Dielectric barrier discharges and nanosecond capillary discharges are conducted to validate the scheme. We find that the propane oxidation processes are primarily controlled by total deposited energy, with minimal dependence on the discharge type (nCD or DBD).
	
\section*{Authors contributions}
Zhenyang Li: Writing-original draft, Formal analysis, Data curation. Bo Yin: Investigation, Experiment development. Qifu Lin: Investigation, Supervision. Yifei Zhu: Writing-review \& editing, Conceptualization. Yun Wu: Funding acquisition, Project administration.

\section{Introduction}

Non-equilibrium plasma-assisted combustion (PAC) offers an effective approach for controlling ignition in internal combustion engines, industrial burners, and aviation engines, thereby enhancing flame stability, reducing pollutant emissions, and expanding ignition limits~\cite{ju2015plasma,ju2016plasma,starikovskiy2013plasma}. Understanding the detailed mechanism of PAC requires studying the kinetic pathways of PAC and quantifying their reaction rates. The kinetics of non-equilibrium plasma-assisted small molecules like methane have been extensively studied~\cite{mao2018effects,mao2019methane,mao2019numerical,lefkowitz2015species,sun2022temperature}.

Propane, as the smallest n-alkane with low-temperature reactivity, serves as a typical molecule for studying the kinetics of large alkanes~\cite{welz2015new,merchant2015understanding}, attracting significant attention from researchers. Adamovich et al.~\cite{adamovich2015kinetic} proposed a plasma-assisted propane-air combustion kinetic model, which showed good agreement with experimental temperature and time-resolved OH density measurements. They further studied the plasma-assisted propane oxidation kinetics in $\rm C_3H_8/O_2/Ar$ mixtures~\cite{eckert2018kinetics}, validated the kinetic model with experiments on alkane small molecule density at different temperatures, and found large discrepancies for $\rm C_2H_2$ and $\rm CH_3CHO$. Chen et al.~\cite{chen2023kinetic} employed molecular beam mass spectrometry with tunable synchrotron vacuum ultraviolet photoionization (SVUV-PIMS) to measure hydrocarbons and oxygenated intermediates during plasma-assisted $\rm C_3H_8/O_2/Ar$ processes, reporting numerous oxygenated intermediates and establishing a kinetic model. However, significant discrepancies existed between the model predictions and experimental measurements for certain hydrocarbons and intermediates. Ban et al.~\cite{ban2023effects} developed a propane-air plasma combustion mechanism and validated it against OH density measurements from experiments~\cite{wu2011plasma} that lacked sufficient validation. Based on the above analysis, current kinetic models still exhibit significant discrepancies with experimental measurements in predicting certain hydrocarbons and oxygenated intermediates. Furthermore, current kinetic models are validated under specific single pulse deposition energies, necessitating further validation under larger deposition energy scales.

To validate the kinetics of the plasma-assisted oxidation process in $\rm C_3H_8/O_2/Ar$ mixtures under large deposition energy scales, stable discharge characteristics and spatiotemporal uniformity are required in experiments. Fast ionization wave (FIW), a nanosecond capillary discharge at moderate pressures, is characterized by spatiotemporal uniformity of discharge, high reduced electric field, high electron energy, and fast propagation speed~\cite{akishev2022slow}. FIW serves as an ideal platform for investigating fundamental plasma processes under various discharge parameters and energy deposition conditions~\cite{klochko2015capillary,chng2019electric,klochko2014experimental,lepikhin2016long,chen2021modeling,akishev2022slow,zhu2020scaling}. Due to its uniformity and high energy deposition, FIW is suitable for investigating plasma-assisted fuel cracking and oxidation. FIW has two discharge types depending on the presence or absence of a metal screen. In the absence of a metal screen (referred to as nanosecond capillary discharge, nCD), the discharge propagation pattern shifts from single-direction to dual-direction propagation, improving discharge channel uniformity. Consequently, nCD is particularly suitable for validating the accuracy of $\rm C_3H_8/O_2/Ar$ plasma combustion mechanisms under varying energy deposition conditions.

In this study, we investigate the mechanism of plasma-assisted low-temperature oxidation of propane using numerical simulations and capillary discharge experiments. Section~\ref{Exp_model} outlines the numerical model, the kinetic schemes, and the experimental setups. The validation of the kinetic schemes against capillary discharge experiments and DBD experiments~\cite{chen2023kinetic} is detailed in Section~\ref{Validation_of_mechanism}. An uncertain analysis of the kinetic schemes is performed using path analysis in Section~\ref{res_DBD}. Section~\ref{res_comparison} compares the differences between nCD and DBD in the process of propane oxidation. Finally, we provide conclusions in Section~\ref{conclusions}.

\section{Experiment and modelling method}\label{Exp_model}

\subsection{The experiment settings}
\begin{figure}[t!h!]
	\centering
	\includegraphics[width=0.9\textwidth]{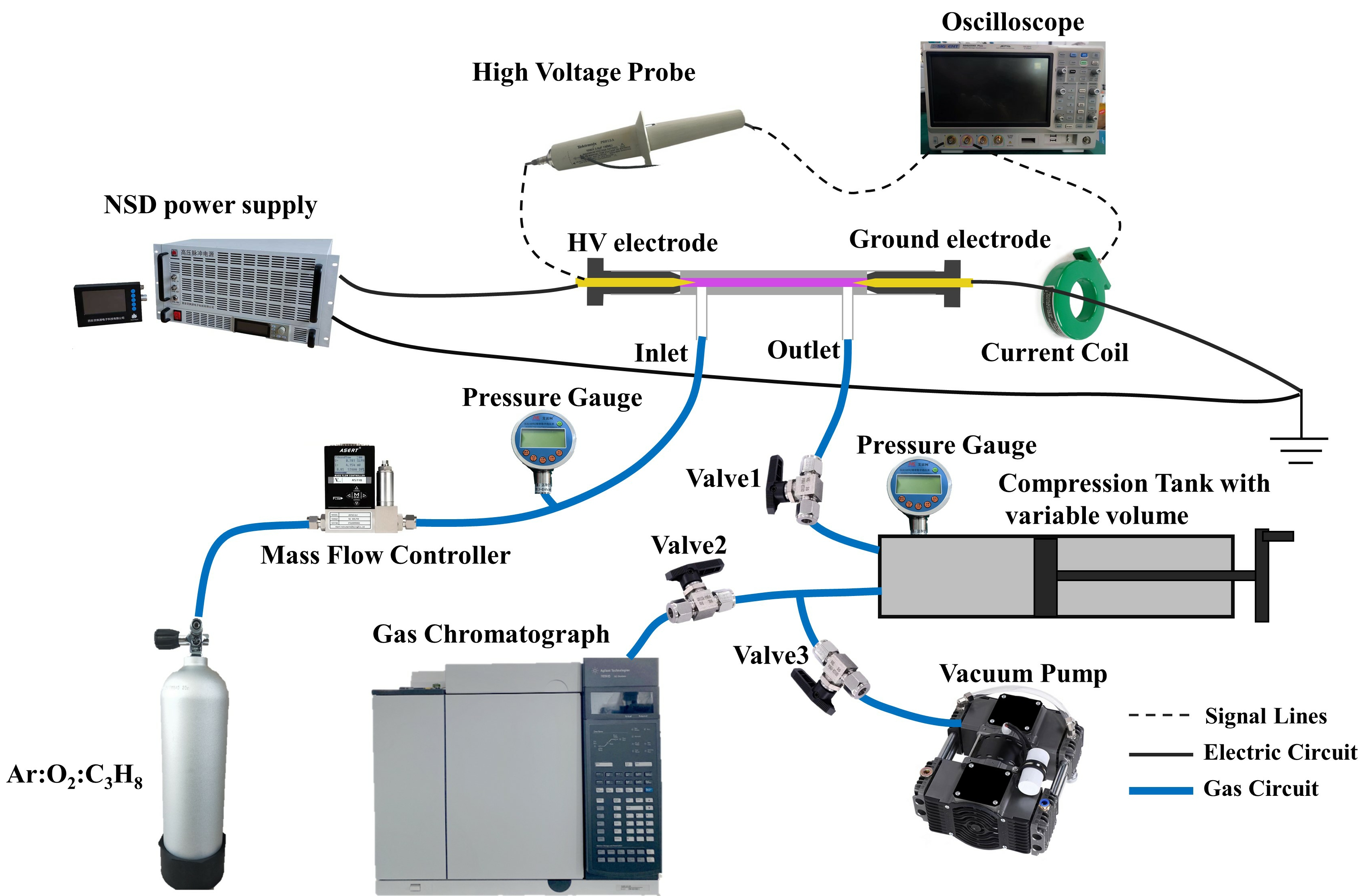}
	\caption{The shematic of the nCD assisted propane oxidation experiment.}
	\label{exp-FIW}
\end{figure}

The experimental setup of the nCD-assisted propane oxidation is depicted in Figure~\ref{exp-FIW}. This setup primarily comprises a gas supply and control system, a plasma discharge system, and a gas collection and analysis system. The discharge system includes a high-voltage electrode, a ground electrode, a quartz tube, and insulating rubber gaskets. The cross-sectional diameter and length of the discharge region are 4.5 mm and 50 mm, and the distance between the two electrodes is 44 mm.

\begin{figure}[t!h!]
	\centering
	\includegraphics[width=0.9\textwidth]{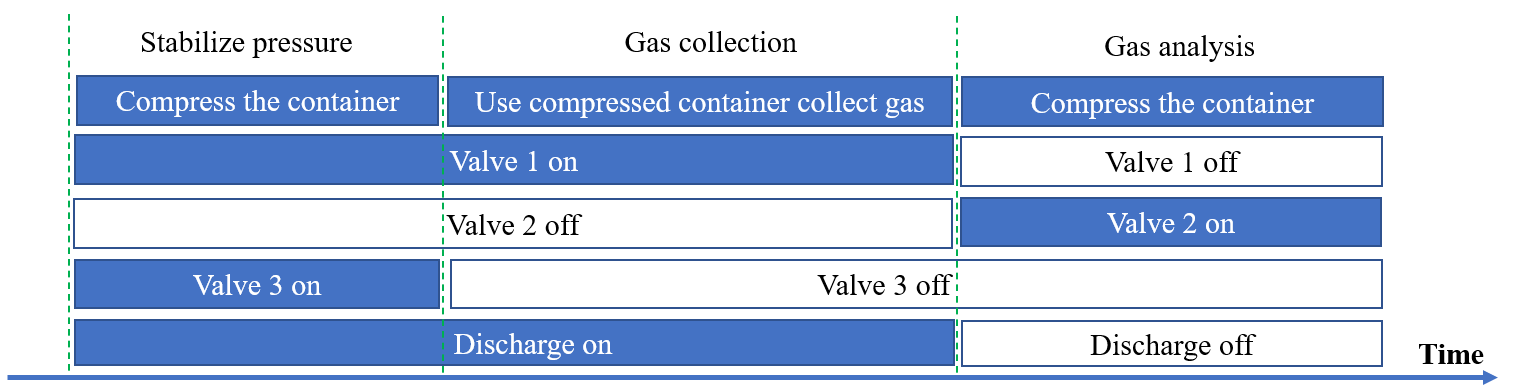}
	\caption{The nCD experimental operation sequence diagram on the process of gas collection.}
	\label{GasCollection}
\end{figure}

The gas mixture in the cylinder consists of Ar, $\rm O_2$, and $\rm C_3H_8$ in a ratio of 92.46:6.33:1.21. The gas flow rate is controlled using a flowmeter (ASERT AST10-ALC 100 sccm) and is set to 30 sccm. Gas flow is controlled using a flowmeter (ASERT AST10-ALC 100 sccm) set at 30 sccm, and pressure is maintained at 3000 Pa by a rotary vane pump, monitored by a vacuum gauge (AILEIKE TT\&C Equipment Co., Ltd., ALKC400HJ, 0-1 atm absolute pressure gauge, 0.05\%FS). As shown in Fig.~\ref{GasCollection},  valve 3 is adjusted to stabilize the gas pressure before collecting the gas. During collection, valves 3 and 2 are closed, and a compression tank with variable volume stabilizes and stores the low-pressure gas, which is later pressurized to atmospheric pressure and analyzed using a gas chromatograph (GC) (Agilent 7890B) for species concentrations.

The discharge is powered by a nanosecond pulse discharge power supply (Xi'an Lingfengyuan Electronic Technology Co., Ltd., HVP-20P), providing continuous pulse outputs for steady-state gas collection. Each pulse has triangular wave shape with 50 ns rise and fall times, operating at a frequency of 1 kHz and a voltage range of 4.2 kV to 5 kV. Discharge voltage is monitored with a high-voltage probe (Tektronix, P6015A), and discharge current is measured with a current coil (Pearson 6585). Both voltage and current data are recorded using an oscilloscope (Siglent SDS2354X 350 MHz).

\subsection{Modelling methods} \label{CPCC-Model}

A new coupled plasma and combustion chemistry solver, CPCC, has been developed by integrating the ZDPlasKin code~\cite{pancheshnyi2008computer} with a custom-built combustion code in F90 format. CPCC supports sensitivity and path flux analysis and interfaces with popular post-processing software such as QtPlaskin~\cite{QTPlaskin} and PumpKin~\cite{markosyan2014pumpkin}. Detailed descriptions of CPCC can be found in~\cite{qiu2023numerical}.The electron impact process is solved using the Boltzmann equation solver BOLSIG+~\cite{hagelaar2005solving}, which is integrated within ZDPlasKin. The plasma kinetic mechanism and ground state kinetic mechanism are alternately solved using the splitting method. For efficient solution of large combustion kinetic mechanisms, CPCC utilizes the semi-implicit ODE solver VODPK~\cite{hindmarsh1983odepack} and the fifth-order implicit Runge-Kutta ODE solver, RADAU5~\cite{2001Radau5}.

The key parameter of the reduced electric field ($E/N$) is derived from experimental current data using the following equation:
\begin{equation}
	 I = n _ { e } \mu ( \frac{E}{N} ) E e S \label{1}
\end{equation}
where $n_e$ is the electron density (initially set to $1.0\times10^9 cm^{-3}$), $I$ is the electron mobility dependent on the reduced electric field, $N$ is the natural particle density of last time, $E$ is the electric field strength, $e$ is the elementary charge, and $S$ is the cross-sectional area of the plasma. $E$ is iteratively obtained using equation~\ref{1} through the Newton iteration method. The intial electron density is given a empirical value, because $E$ is insensitive to the intial electron density, which will iterate consistently to a reasonable value in the first few time steps without affecting the result of the calculation. Then $E/N$ is obtained by dividing $E$ by $N$. 

In present model, as the Poisson equation is not solved, the deposition energy is assumed to be an adjustable parameter and controls the discharge duration~\cite{mao2019methane}. For DBD experiment, the present model's deposition energy stands at $1.85 \times 10^{-5} cm^{-3} $, significantly lower than the experimentally measured value. Furthermore, the deposition energy control is not implemented. For nCD experiment, To ensure good agreement with the experimental results, the model utilized 40\% of the experimental deposited energy to regulate the discharge time.

To account for primary energy loss due to conduction to the quartz channel walls, similar to the approach in Ref.~\cite{lefkowitz2015species}, the model includes surface heat transfer between the plasma and the quartz wall. In the energy equation for the nCD experiment, an additional energy loss term is introduced:
  \begin{equation}
	\rho C _ { v } \frac { \mathrm { d } T _ { \mathrm { gas } } } { \mathrm { d } t } = - \sum _ { i = 1 } ^ { i _ { \max } } H _ { i } \omega _ { i } - \frac { T _ { \mathrm { gas } } - T _ { \mathrm { wall } } } { \left( \frac { 1 } { A _ { 1 } h _ { 1 } } + \frac { \ln \left( r _ { 2 } / r _ { 1 } \right) } { 2 \pi \lambda L } \right) V _ { p } }\label{2}
\end{equation}
where $\rho$ is the gas density, $C_v$ is the specific heat capacity at constant volume of the gas, $H_i$ and $\omega _i$ represent the enthalpy of formation and production/loss rate of component i, respectively. $T_{wall}$ is the outer wall temperature of the quartz tube (set to 300K), $A_1$ is the contact area between the plasma and the inner wall of the quartz tube, $h_1$ is the heat transfer coefficient between the plasma and the inner wall of the quartz tube (which has been adjusted to 10 $W/(m^2 K)$ based on experimental results due to the cylindrical shape and small diameter of this structure, making empirical formulas unsuitable for estimating the heat transfer coefficient), $r_2$ and $r_1$ are the outer and inner radius of the quartz tube, $\lambda$ is the thermal conductivity of quartz (which is 1.09 $W/(m K)$), $L$ is the length of the discharge region, and $V_p$ is the volume of the plasma.

For a DBD configuration where the discharge region is rectangular, the energy loss term in the gas energy equation is adjusted as follows:
\begin{equation}
	\rho C _ { v } \frac { \mathrm { d } T _ { \mathrm { gas } } } { \mathrm { d } t } = - \sum _ { i = 1 } ^ { i _ { \max } } H _ { i } \omega _ { i } - \frac { T _ { \mathrm { gas } } - T _ { \mathrm { wall } } } { \left( \frac { 1 } { A _ { 2 } h _ { 2 } } + \frac { \delta } { A _ { 2 } \lambda } \right) V _ { p } }\label{3}
\end{equation}
\begin{equation}
	h _ { 2 } = \frac { \lambda } { L N_u }\label{4}
\end{equation}
\begin{equation}
	N_u = 0.664  R_e ^ { 1 / 2 } P_r  ^ { 1 / 3 }\label{5}
\end{equation}
\begin{equation}
	 P_r  = \frac { C _ { p } \mu } { \lambda }\label{6}
\end{equation}
\begin{equation}
	 R_e  = \frac { \rho v L } { \mu }\label{7}
\end{equation}
where $A_2$ is the contact area between the plasma and the upper and lower walls, $h_2$ is the surface heat transfer coefficient of the walls (calculated using equation~\ref{4}), and the Nusselt number $N_u$ is calculated using equation~\ref{5} from the book~\cite{HeatTransferTheory}. $P_r$ is the Prandtl number (calculated by equation~\ref{6}), $R_e$ is the Reynolds number (calculated by equation~\ref{7}), $\mu$ is the dynamic viscosity, $v$ is the airflow velocity, and $C_p$ is the specific heat capacity at constant pressure of the gas.

This kinetic model includes both plasma chemistry and classical combustion chemistry. The combustion chemistry model incorporates a total of 2521 reactions, which have been updated based on the scheme proposed in Ref.~\cite{chen2023kinetic}. Specific modifications derived from experimental data are outlined in Table~\ref{UpdatedReaction}. In addition, several combustion reactions have been added and replaced, drawing from various established mechanisms such as HP-Mech v3.3~\cite{reuter2018counterflow}, AramoMech3.0~\cite{zhou2018experimental}, NUIGMech1.1~\cite{ramalingam2021chemical} (see supplementary material of Table S1). The plasma chemistry model consists of 2202 reactions, involving vibrationally excited species such as $\rm C_3H_8(v1)$-$\rm C_3H_8(v27)$, $\rm CH_4(v2,4)$, $\rm CH_4(v1,3)$, $\rm O_2(v1)$-$\rm O_2(v4)$; electronically excited species like $\rm O_2 (a^1 \Delta _g )$, $\rm O_2 (b^1 \Sigma_g^+ )$, $\rm O_2(4.5eV)$(summation of $\rm O_2 \left( c ^1 \Sigma _ {u} ^{ -} \right)$, $\rm O_2 \left( C ^3 \Delta  _ {u} \right)$, $\rm O_2 \left( A ^3 \Sigma _ {u} ^{ +} \right)$), $\rm O(^1D)$, $\rm O(^1S)$, $\rm Ar(1s_2)$, $\rm Ar(1s_4)$, $\rm Ar(11.55eV)$, $\rm Ar(2p_1)$; ions like $\rm O^+$, $\rm O_2^+$, $\rm O^-$, $\rm O_2^-$, $\rm O_3^-$, $\rm CH_4^+$, $\rm CH_3^+$, $\rm CH_2^+$, $\rm CH_5O^+$, $\rm C_2H_3O^+$, $\rm CHO_3^-$, $\rm CO_3^-$, $\rm CHO_2^-$, $\rm OH^-$, $\rm C_2H_4^+$, $\rm C_2H_5^+$, $\rm C_2H^+$, $\rm C_2H_3^+$, $\rm H_5C_2O^+$, $\rm C_3H_7^+$, $\rm C_4H_9^+$, $\rm C_3H_8^+$, and electron. The plasma chemistry model includes chemical reactions involving vibrationally excited states, vibrational relaxation reactions, charge transfer reactions, and reaction channels involving $\rm O(^1D)$, $\rm O(^1S)$, $\rm O_2 (a^1 \Delta _g )$ , $\rm O_2 (b^1 \Sigma_g^+ )$, $\rm O_2 (4.5eV)$, $\rm Ar(1s_2)$, $\rm Ar(1s_4)$, $\rm Ar(11.55eV)$, $\rm Ar(2p_1)$ with fuel fragments. The electron collision dissociation cross-section for $\rm C_3H_8$ is sourced from Ref.~\cite{janev2002collision}, while reactions involving electron collision dissociation for $\rm CH_4$, including dissociation and ionization, are adopted from Ref.~\cite{mao2019numerical}.  The plasma chemistry of Ar and $\rm O_2$ are taken from Ref.~\cite{zhu2018optical}.

\begin{table}[]
	\footnotesize
	
	\caption{Updated plasma reactions and combustion reactions in this work.}\label{UpdatedReaction}
	\begin{threeparttable} 
	\scalebox{0.82}{	
	\begin{tabular}{cccc}
	\hline
	\multicolumn{1}{l}{}&\multicolumn{1}{l}{Plasma reaction}                                   & \multicolumn{1}{l}{\begin{tabular}[c]{@{}l@{}}Original rate \\ constant\end{tabular}} & \multicolumn{1}{l}{\begin{tabular}[c]{@{}l@{}}Updated rate \\ constant\end{tabular}} \\ \hline
	R1&\multicolumn{1}{l}{$\rm e+C_3H_8\rightarrow e+H_2+C_3H_6$}      & \multicolumn{1}{l}{$f$($\sigma_1$)\cite{janev2002collision}} & \multicolumn{1}{l}{$0.8\times f$($\sigma_1$)}\\
	R2&\multicolumn{1}{l}{$\rm e+C_3H_8\rightarrow e+CH_2+C_2H_6$}     & \multicolumn{1}{l}{$f$($\sigma_2$)\cite{janev2002collision}} & \multicolumn{1}{l}{$12\times f$($ \sigma_2$)}\\
	R3&\multicolumn{1}{l}{$\rm e+C_3H_8\rightarrow e+CH_4+C_2H_4$}     & \multicolumn{1}{l}{$f$($\sigma_3$)\cite{janev2002collision}} & \multicolumn{1}{l}{$0.36\times f$($\sigma_3$)} \\
	R4&\multicolumn{1}{l}{$\rm e+C_3H_8\rightarrow e+H_2+H_2+C_3H_4$}   & \multicolumn{1}{l}{$f$($\sigma_4$)\cite{janev2002collision}} & \multicolumn{1}{l}{$0.75\times f$($\sigma_4$)} \\
	R5&\multicolumn{1}{l}{$\rm e+C_3H_8\rightarrow e+CH_2+C_2H_2+2H_2$} & \multicolumn{1}{l}{0.0}  & \multicolumn{1}{l}{$21\times f$($\sigma_2$)} \\ \hline
	&\multicolumn{1}{l}{Combustion reaction}                               & \multicolumn{1}{l}{\begin{tabular}[c]{@{}l@{}}Original Arrhenius \\ equation of A\end{tabular}} & \multicolumn{1}{l}{\begin{tabular}[c]{@{}l@{}}Updated Arrhenius \\ equation of A\end{tabular}} \\ \hline
	R6&\multicolumn{1}{l}{$\rm CH_3OH+O_2\leftrightarrow CH_3O+HO_2$}        & \multicolumn{1}{l}{$3.58\times 10^4$}    & \multicolumn{1}{l}{0.0~\cite{klippenstein2011uncertainty}}  \\
	R7&\multicolumn{1}{l}{$\rm n$-$\rm C_3H_7O_2+HO_2\leftrightarrow n$-$\rm C_3H_7O_2H+O_2$}  & \multicolumn{1}{l}{$1.75\times 10^{10}$}   & \multicolumn{1}{l}{$6.0\times 10^8$}  \\
	R8&\multicolumn{1}{l}{$\rm i$-$\rm C_3H_7O_2+HO_2\leftrightarrow i$-$\rm C_3H_7O_2H+O_2$}  & \multicolumn{1}{l}{$1.75\times 10^{10}$}   & \multicolumn{1}{l}{$6.0\times 10^8$}\\
	R9&\multicolumn{1}{l}{$\rm i$-$\rm C_3H_7O_2+C_2H_5O_2\leftrightarrow CH_3COCH_3+C_2H_5OH+O_2$}   & \multicolumn{1}{l}{0.0}       & \multicolumn{1}{l}{$5.0\times 10^{16}$}  \\
	R10&\multicolumn{1}{l}{$\rm n$-$\rm C_3H_7O_2+n$-$\rm C_3H_7O_2\leftrightarrow C_2H_5CHO+n$-$\rm C_3H_7OH+O_2$}  & \multicolumn{1}{l}{$1.42\times 10^{16}$}          & \multicolumn{1}{l}{$7.0\times 10^{15}$}\\
	R11&\multicolumn{1}{l}{$\rm n$-$\rm C_3H_7O_2+i$-$\rm C_3H_7O_2\leftrightarrow C_2H_5CHO+i$-$\rm C_3H_7OH+O_2$}  & \multicolumn{1}{l}{$1.409\times 10^{16}$}        & \multicolumn{1}{l}{$4.109\times 10^{16}$} \\
	R12&\multicolumn{1}{l}{$\rm i$-$\rm C_3H_7O_2+n$-$\rm C_3H_7O_2\leftrightarrow CH_3COCH_3+n$-$\rm C_3H_7OH+O_2$} & \multicolumn{1}{l}{$1.409\times 10^{16}$}           & \multicolumn{1}{l}{$1.169\times 10^{15}$}\\
	R13&\multicolumn{1}{l}{$\rm i$-$\rm C_3H_7O_2+i$-$\rm C_3H_7O_2\leftrightarrow CH_3COCH_3+i$-$\rm C_3H_7OH+O_2$} & \multicolumn{1}{l}{$1.409\times 10^{16}$}          & \multicolumn{1}{l}{$2.109\times 10^{16}$} \\ \hline
	\end{tabular}
	}
	\begin{tablenotes}    
        \footnotesize               
        \item[*] $\sigma_x$ indicate the cross sections for the electron-impact reaction. The rate constant values are calculated by BOLSIG+ (indicated by $f$ ) given E/N and the cross section data. The unit of cross section data is $m^2$. The unit of arrhenius equation pre-exponential factor A is $cm^3 \: mol^{-1}\:  s^{-1}$.          
    \end{tablenotes}            
	
    \end{threeparttable}       
\end{table}

The original model in Section~\ref{Validation_of_mechanism} and Section~\ref{sec_discussion} is defined as the plasma chemistry and classical combustion chemistry in which the reaction rate coefficients are from literature without any modifications. Compared to the model of Ref.~\cite{chen2023kinetic}, the original model reconstruct the plasma chemistry and has the same combustion chemistry. As some reactions are not validated with experiments or quantum chemical calculations, the original model predictions for some species have large uncertainties. So the electron impact cross sections of propane, branch ratios of self-/cross-reactions for $\rm i$-/$\rm n$-$\rm C_3H_7O_2$ and H-abstraction of $\rm i$-/$\rm n$-$\rm C_3H_7O_2$ are fitted within uncertainties according to experimental data and path flux analysis (named as updated model in Section~\ref{Validation_of_mechanism} and Section~\ref{sec_discussion}). The updated reactions are listed in Table~\ref{UpdatedReaction}. Detailed descriptions and uncertainty analysis can be found in Section~\ref{res_DBD}. The species overall predictions of the updated model are highly improved compared to the original model in DBD experiment~\cite{chen2023kinetic} and nCD experiment conditions. Even so, the quantum chemical calculations or fundamental kinetic experiment are still requred to improve the model's prediction capacity.

\section{Validation of the mechanism}\label{Validation_of_mechanism}
\subsection{Validation I:DBD assisted propane oxidation}

\begin{figure}[t!h!]
	\centering
	\includegraphics[width=0.9\textwidth]{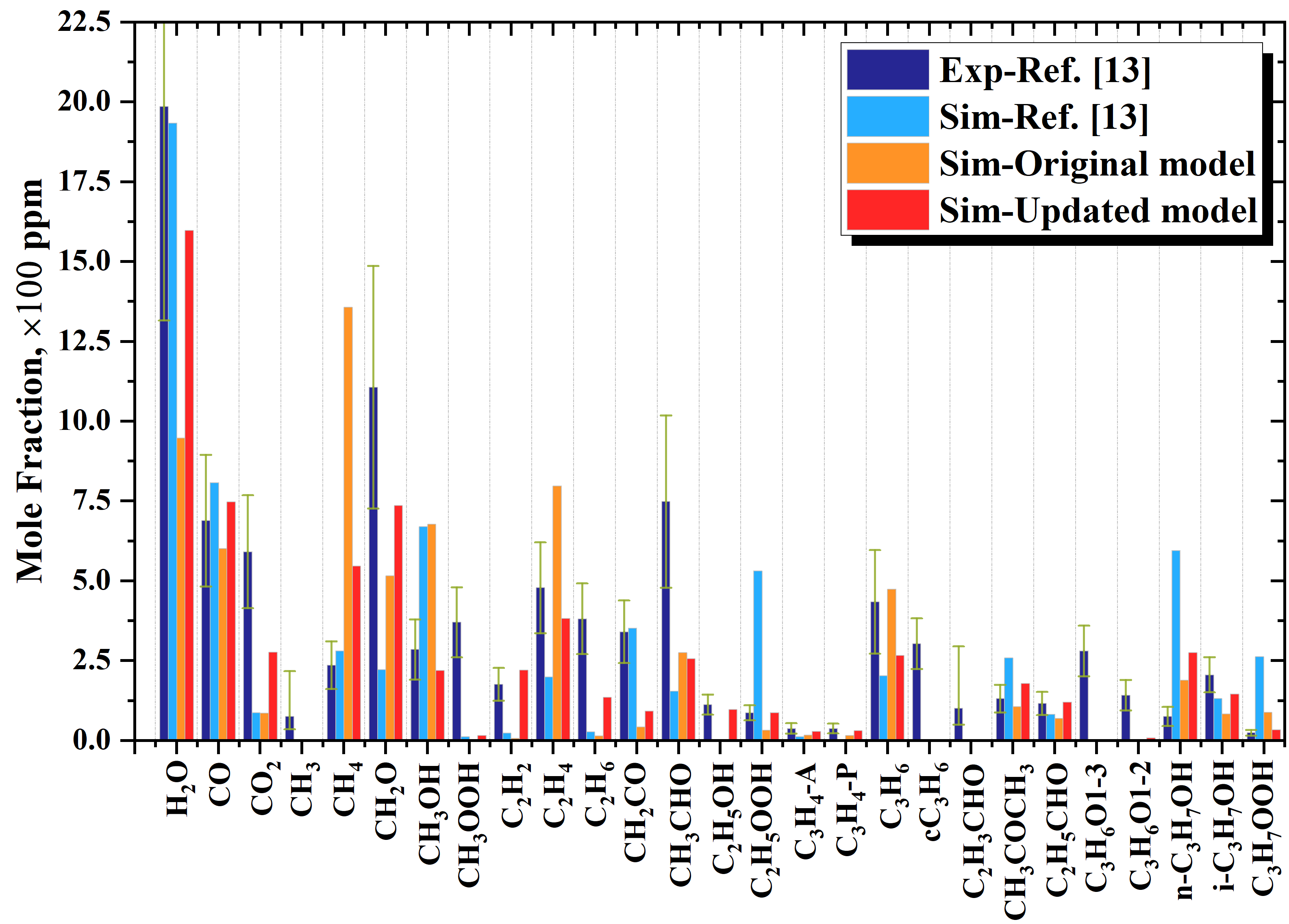}
	\caption{Comparison between measurement and predictions from different kinetics schemes species in repetitively-pulsed nanosecond DBD. The deep blue and light blue bars represent experiment measurement and model prediction results from the Ref.~\cite{chen2023kinetic}.}
	\label{Validation-exp-DBD}
\end{figure}

Model validation is performed by comparing the model predicted and experimental molar fractions of speices in a repetitively-pulsed nanosecond DBD~\cite{chen2023kinetic}, as shown in Fig.~\ref{Validation-exp-DBD}. The experimental conditions involve a deposition energy of $1.64\times 10^{-4} J/cm^3$, at 30 Torr and a discharge region temperature of 340 K, with a $\rm C_3H_8/O_2/Ar$ mixture (mole fraction 4:20:76). The updated model's predictions for several species, including $\rm H_2O$, CO, $\rm CH_2O$, $\rm CH_3OH$, $\rm C_2H_2$, $\rm C_2H_4$, $\rm C_2H_5OH$, $\rm C_2H_5OOH$, $\rm C_3H_4$-$\rm A$, $\rm C_3H_4$-$\rm P$, $\rm C_3H_6$, $\rm CH_3COCH_3$, $\rm C_2H_5CHO$, $\rm i$-$\rm C_3H_7OH$, and $\rm C_3H_7OOH$ are found to be in good agreement with experimental data, falling within the range of experimental measurement errors. The updated model's predictions for species $\rm CO_2$, $\rm CH_4$, $\rm C_2H_6$, $\rm CH_2CO$, $\rm CH_3CHO$, and $\rm n$-$\rm C_3H_7OH$ are within 5 times of the experimental measurement. However, the updated model's predictions for species $\rm CH_3$, $\rm CH_3OOH$, $\rm cC_3H_6$, $\rm C_2H_3CHO$, $\rm C_3H_6O_{1-3}$, and $\rm C_3H_6O_{1-2}$ are over an order of magnitude off. The updated model demonstrates noticeable improvements in prediction accuracy for many species compared to the original model and a previously published model~\cite{chen2023kinetic}.

The reasons for the lower predictions of $\rm C_3H_6O_{1\text{-}2}$ and $\rm C_3H_6O_{1\text{-}3}$ in the updated model are as follows: In Ref.~\cite{bedjanian2017reaction,leonori2015experimental}, $\rm C_3H_6$ is consumed by an O atom to form $\rm C_3H_6O_{1\text{-}2}$ and $\rm C_3H_6O_{1\text{-}3}$ intermediates through the addition of an O atom to the C=C bonds in the center and end carbon atoms of $\rm C_3H_6$. Then, $\rm C_3H_6O_{1\text{-}2}$ and $\rm C_3H_6O_{1\text{-}3}$ intermediates decompose into $\rm CH_3$, $\rm C_2H_5$, $\rm CH_2O$, and H, among others. The reactions provided in Ref.~\cite{bedjanian2017reaction} do not consider these intermediate products ($\rm C_3H_6O_{1\text{-}2}$ and $\rm C_3H_6O_{1\text{-}3}$) but instead directly provide the reactions leading to the final products. This ultimately results in the underprediction of $\rm C_3H_6O_{1\text{-}2}$ and $\rm C_3H_6O_{1\text{-}3}$.

\subsection{Validation II: nCD assisted propane oxidation}

\begin{figure}[t!h!]
	\centering
	\includegraphics[width=0.9\textwidth]{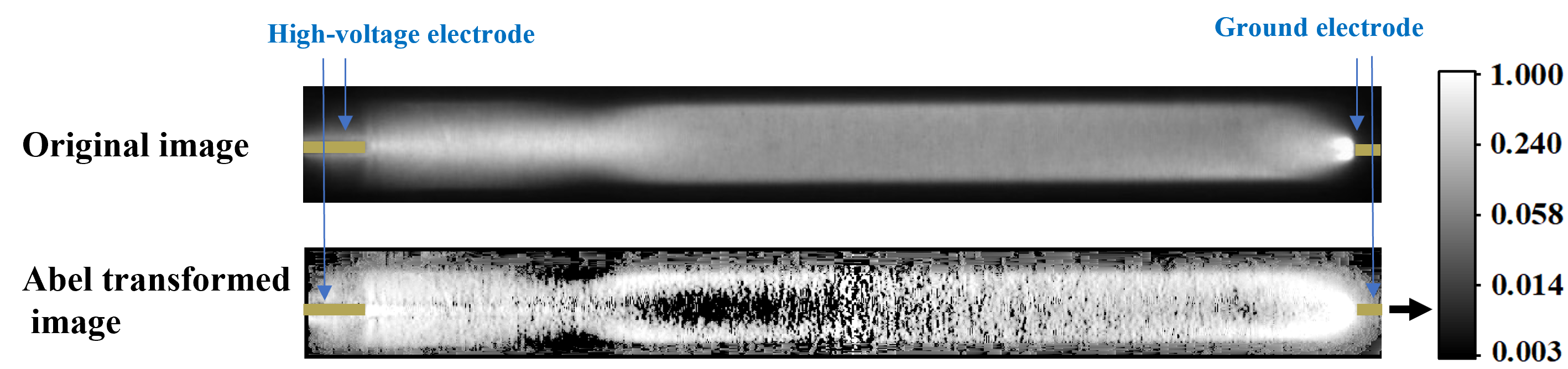}
	\caption{The nCD image captured by high-speed ICCD camera and abel transformed image with voltage of 5.0 kV. The gate width is set to 100ns.}
	\label{exp-FIW-figure}
\end{figure}

The DBD experiment was conducted under conditions with a deposition energy of $1.64\times 10^{-4} J/cm^3$. To further validate the accuracy of the mechanism, nCD experiments were conducted with deposition energies ranging from $1.7\times 10^{-4} J/cm^3$  to $1.53\times 10^{-3} J/cm^3$. The uniformity of the nCD was evaluated using images captured by a high-speed ICCD camera operated at 5.0 kV, which were processed using an Abel transformator, as depicted in Fig.~\ref{exp-FIW-figure}. Due to the enhanced electric field near the electrode tip, localized discharge enhancement was observed on the ground electrode. Conversely, a phenomenon of discharge weakening was noted at distances ranging from 9 to 20 mm away from the high-voltage electrode. These localized effects were considered negligible in comparison to the overall discharge characteristics. Overall, the nCD was found to exhibit relatively uniform behavior across the tested conditions.

\begin{figure}[t!h!]
	\centering
	\begin{subfigure}{0.475\linewidth}
		\centering
		\includegraphics[width=1.0\textwidth]{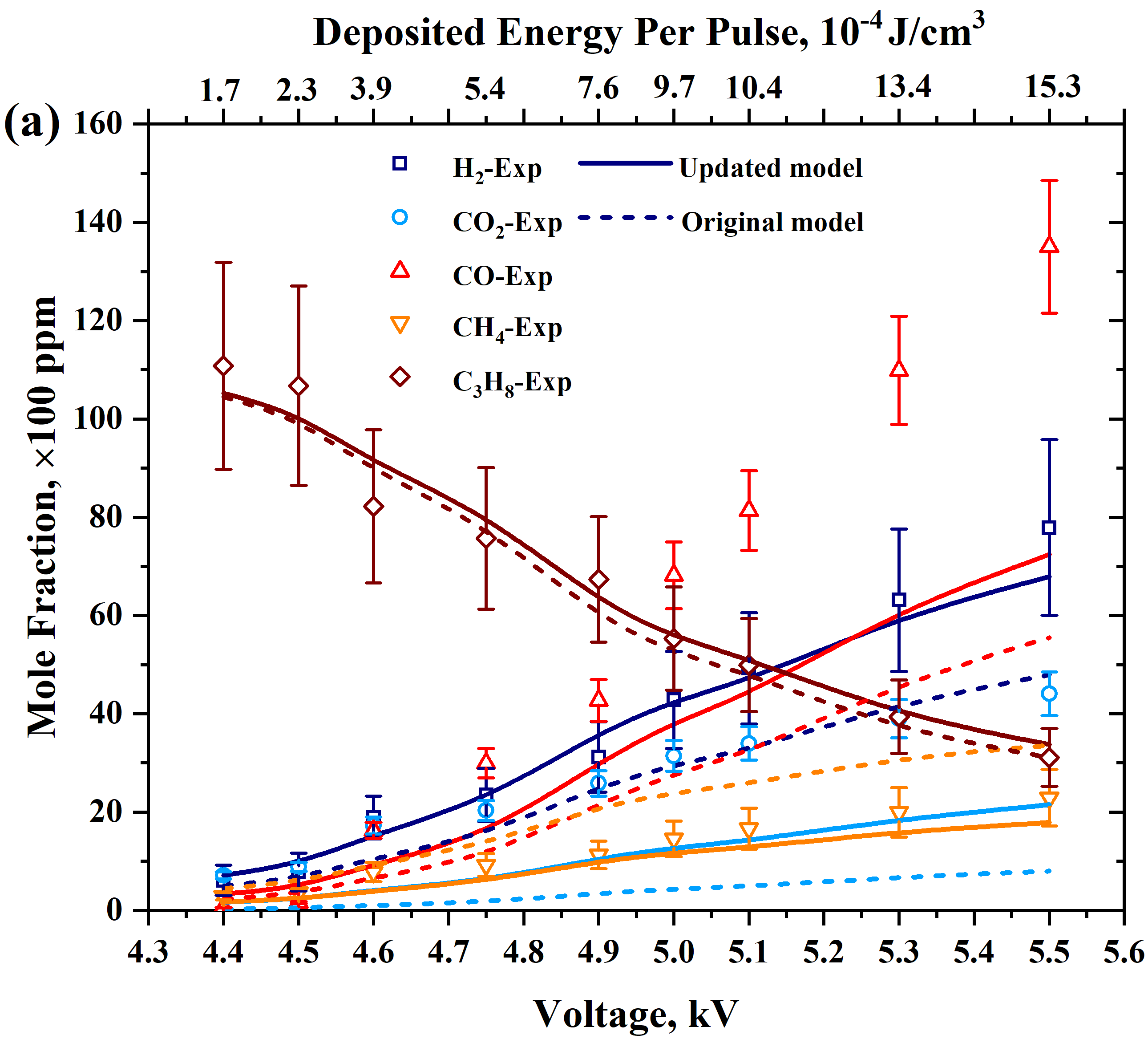}
    \end{subfigure}
	\begin{subfigure}{0.46\linewidth}
		\centering
		\includegraphics[width=1.0\textwidth]{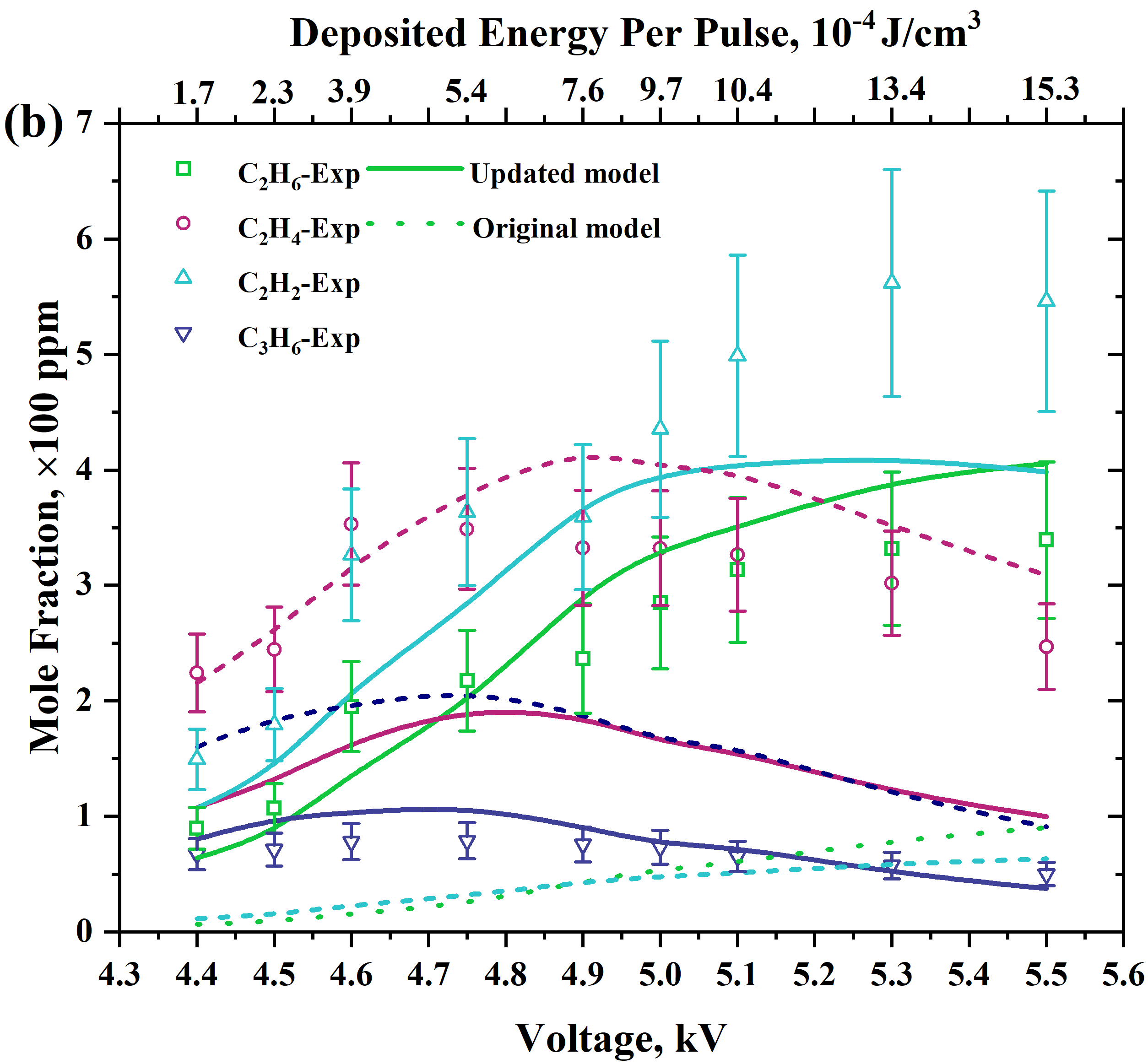}
	\end{subfigure}
	\caption{Comparison between measurement and predictions from different kinetics
	scheme in nCD at different voltages.}
	\label{Validation_With_FIW}
\end{figure}

Figure~\ref{Validation_With_FIW} presents a comparison between experimental measurements and predictions from different kinetic schemes in the nCD conditions. The original model exhibits poor predictions for species such as CO, $\rm H_2$, $\rm CO_2$, $\rm CH_4$, $\rm C_2H_6$, $\rm C_2H_2$, $\rm C_3H_6$. In contrast, the updated model shows significant improvements across these species. However, despite these enhancements, notable discrepancies remain in the predicted mole fractions of $\rm CO_2$, $\rm C_2H_4$, and CO, which are under-predicted by 55\%, 45\%, and 45\%, respectively.

The comparison between measurement and predictions from different kinetics scheme in nCD is shown in Fig.~\ref{Validation_With_FIW}. The original model has poor predictions in species CO, $\rm H_2$, $\rm CO_2$, $\rm CH_4$, $\rm C_2H_6$, $\rm C_2H_2$, $\rm C_3H_6$. The updated model has great improvement in above mentioned species. However, compared to experimental measurements, the updated model still has significant differences in the mole fractions of $\rm CO_2$, $\rm C_2H_4$, and CO. The mole fractions of updated model are under-predicted by 55\%, 45\%, and 45\%, respectively. The predictions of remaining species are generally well. 

The updated model's predictions for $\rm C_3H_8$, $\rm H_2$, $\rm CH_4$, and $\rm C_2H_6$ generally fall within the experimental error range. For $\rm C_3H_6$, while it is overpredicted by 30\% at low deposition energy, the agreement improves at higher deposition energies. The trend of $\rm C_2H_2$ predicted by the updated model aligns with experimental data, albeit with molar fractions approximately 30\% lower than measured values. This suggests that while the added reaction pathways in the updated model are correct, further research is needed to refine the cross-sectional data used.

\section{Results and discussion}\label{sec_discussion}

\subsection{Uncertain analysis in the plasma and combustion kinetic mechanism}\label{res_DBD} 

The updated model was derived from the original model through adjustments to several reactions based on experimental findings and pathway flux analysis. Specifically, the modifications listed in Table~\ref{UpdatedReaction} are discussed in detail as follows.
\begin{figure}[htbp]
	\centering
	\includegraphics[width=1.0\linewidth]{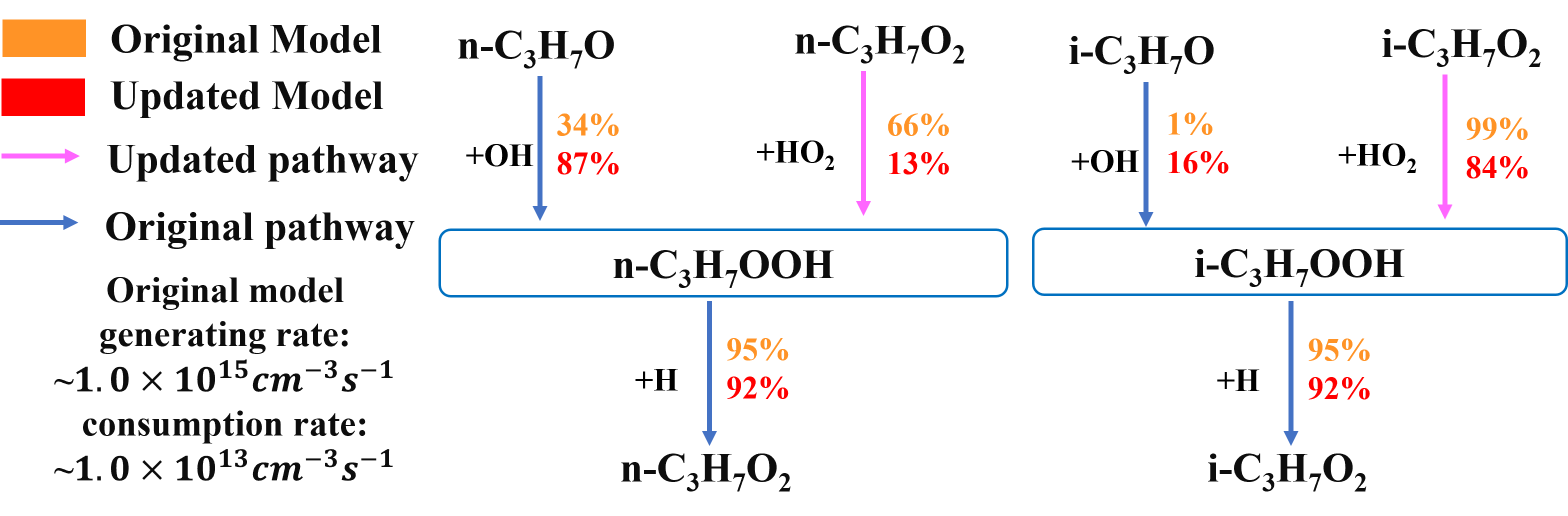}
	\caption{The pathway analysis of $\rm i\text{-}/n\text{-}C_3H_7OOH$ in the updated model and the original model on the process of DBD assisted propane oxidation.}
	\label{pathway_of_C3H7OOH}
\end{figure}

In the original model, significant over-predictions of $\rm i\text{-}/n\text{-}C_3H_7OOH$ were observed during the DBD-assisted propane oxidation process, as depicted in Fig.~\ref{Validation-exp-DBD}. The generation rate of $\rm i\text{-}/n\text{-}C_3H_7OOH$ was found to be two orders of magnitude larger than its consumption rate. Pathway flux analysis based on the original model identified that the primary generation pathways for $\rm i\text{-}/n\text{-}C_3H_7OOH$ involved reactions $\rm i\text{-}/n\text{-}C_3H_7O_2$ with $\rm HO_2$ (reactions R7 and R8, respectively), as illustrated in Fig.~\ref{pathway_of_C3H7OOH}. However, the rate coefficients for R7 and R8 used in the original model were adapted from analogous reactions such as those between $\rm CH_3O_2$ and $\rm HO_2$ reported in Ref.~\cite{tsang1987chemical}. These rate coefficients are valid within a temperature range below 338 K and are subject to an error of up to 5. Given the temperature range of 340-500 K in the current study and the inherent uncertainties associated with the analogy method, significant uncertainties were anticipated for R7 and R8. Therefore, adjustments were made to the rate coefficients of R7 and R8, reducing them by a factor of 0.034 based on insights from the DBD experiment~\cite{chen2023kinetic}. The updated model now indicates that 87\% of $\rm n\text{-}C_3H_7OOH$ is generated from the reaction of $\rm n\text{-}C_3H_7O$ with OH, while 84\% of $\rm i\text{-}C_3H_7OOH$ is produced via reaction R8.

\begin{figure}[htbp]
	\centering
	\includegraphics[width=1.0\linewidth]{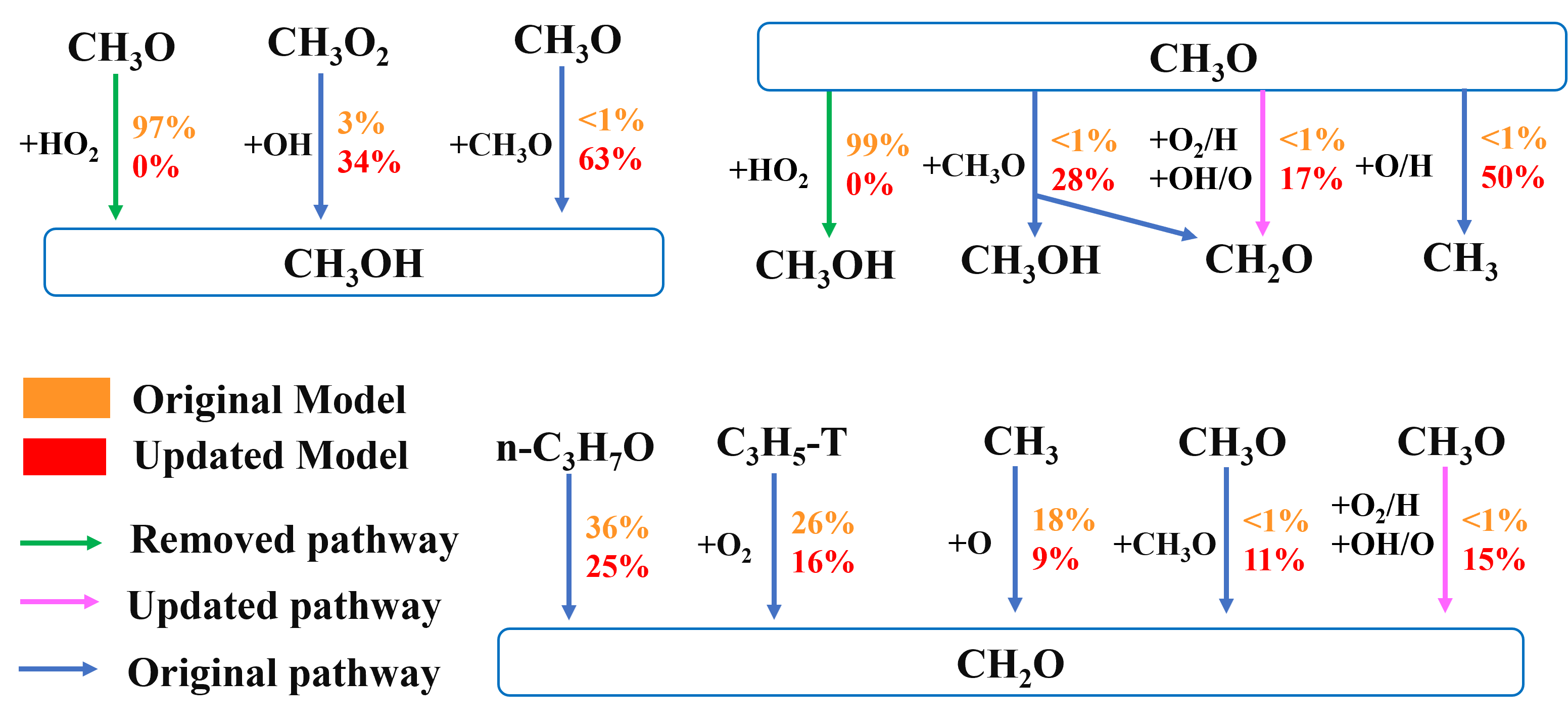}
	\caption{The pathway analysis of $\rm CH_3OH$, $\rm CH_3O$ and $\rm CH_2O$ in the updated model and the original model on the process of DBD assisted propane oxidation.}
	\label{pathway_of_CH3OH_CH2O}
\end{figure}

Both the model by Chen et al~\cite{chen2023kinetic} and the original model exhibit under-prediction of $\rm CH_2O$ and over-prediction of $\rm CH_3OH$. As depicted in Fig.~\ref{pathway_of_CH3OH_CH2O}, pathway flux analysis of the original model indicates that 97\% of $\rm CH_3OH$ is generated via the reaction of $\rm CH_3O$ with $\rm HO_2$ (R6). However, it was noted that the reaction rate coefficients of R6 were sourced from Ref.~\cite{klippenstein2011uncertainty}, which states that the reaction between $\rm CH_3OH$ and $\rm O_2$ produces $\rm CH_2OH$ and $\rm HO_2$, not $\rm CH_3O$ and $\rm HO_2$. Consequently, in the updated model, reaction R6 has been removed. The updated model's prediction for $\rm CH_3OH$ now aligns well with experimental measurements, with 63\% of $\rm CH_3OH$ being produced by the self-reaction of $\rm CH_3O$ (Reaction~\ref{R-1}), and 34\% being generated through Reaction~\ref{R-2}.

\begin{equation}
	\rm CH_3O+CH_3O\leftrightarrow CH_3OH+CH_2O \label{R-1}
\end{equation}
\begin{equation}
	\rm CH_3O_2+OH\leftrightarrow CH_3OH+O_2 \label{R-2}
\end{equation}

To address the lack of certain $\rm CH_3O$ consumption pathways, the $\rm CH_3O$ sub-mechanism was supplemented by referencing the HP-Mech v3.3 mechanism~\cite{reuter2018counterflow} (as shown in Table~\ref{Combustion_reactions}). The removal of R6 and the supplement of $\rm CH_3O$ consumption reactions led to significant changes in the $\rm CH_3O$ consumption pathway, as illustrated in Fig.~\ref{pathway_of_CH3OH_CH2O}. In the original model, $\rm CH_3O$ was almost exclusively consumed by the reverse reaction of R6. In contrast, in the updated model, half of $\rm CH_3O$ is consumed to form $\rm CH_3$, while the remaining $\rm CH_3O$ is consumed to form $\rm CH_2O$ and $ \rm CH_2O$. Consequently, the prediction of $\rm CH_2O$ in the updated model has improved and now aligns well with experimental measurements due to the removal of R6 and the supplement of $\rm CH_3O$ consumption reactions.

\begin{equation}
	\rm n\text{-}C_3H_7O\rightarrow C_2H_5+CH_2O \label{R-3}
\end{equation}
\begin{equation}
	\mathrm { C } _ { 3 } \mathrm { H } _ { 5 } \text{-} \mathrm { T } + \mathrm { O } _ { 2 } \leftrightarrow \mathrm { CH } _ { 3 } \mathrm { CO } + \mathrm { CH } _ { 2 } \mathrm { O }\label{R-4}
\end{equation}
\begin{equation}
	\mathrm { CH } _ { 3 } + \mathrm { O } \leftrightarrow \mathrm { CH } _ { 2 } \mathrm { O } + \mathrm { H }\label{R-5}
\end{equation}
\begin{equation}
	\rm CH_3O+O_2/H/OH/O \leftrightarrow CH_2O+HO_2/H_2/H_2O/OH \label{R-6}
\end{equation}

\begin{figure}[htbp]
	\centering
	\includegraphics[width=1.0\linewidth]{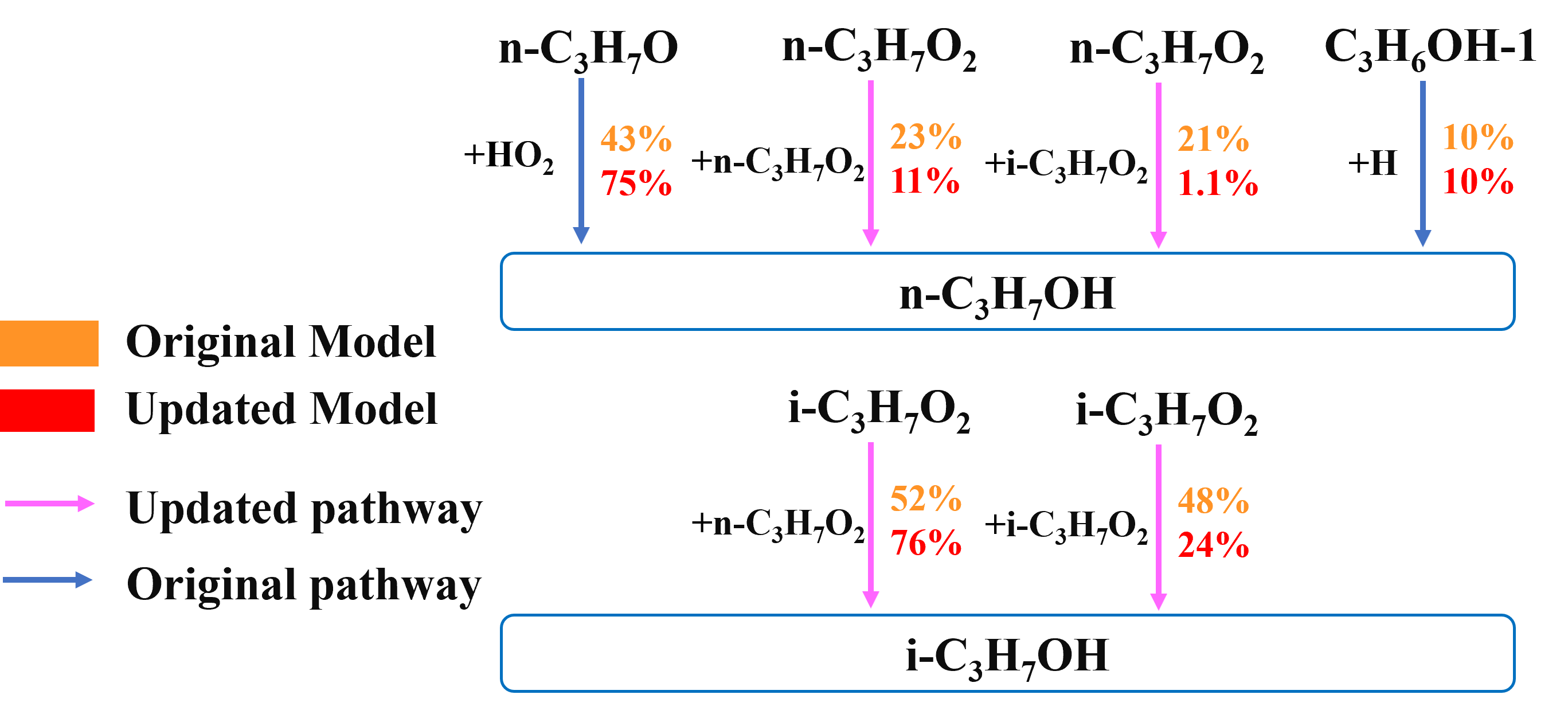}
	\caption{The pathway analysis of $\rm n\text{-}C_3H_7OH$ and $\rm i\text{-}C_3H_7OH$ in the updated model and the original model on the process of DBD assisted propane oxidation.}
	\label{pathway_of_C3H7OH}
\end{figure}

As depicted in Fig.~\ref{Validation-exp-DBD}, the original model exhibited over-predictions of $\rm n\text{-}C_3H_7OH$ and under-predictions of $\rm i\text{-}C_3H_7OH$. It was observed that nearly all $\rm i\text{-}C_3H_7OH$ and almost half of $\rm n\text{-}C_3H_7OH$ were produced through self- and cross-reactions of $\rm n\text{-}/i\text{-}C_3H_7O_2$ (Reactions R10-R13), as illustrated in Fig.~\ref{pathway_of_C3H7OH}. The inaccuracies in the branching ratios of these reactions likely contributed to the over-prediction of $\rm n\text{-}C_3H_7OH$ and the under-prediction of $\rm i\text{-}C_3H_7OH$. The rate coefficients for Reactions R10-R13 were originally derived from Ref.~\cite{welz2015new}. However, the temperature-dependent rate constants used in Ref.~\cite{welz2015new}, sourced from Ref.~\cite{atkinson2006evaluated}, lacked detailed pressure and temperature dependencies. Moreover, the products of Reactions R10-R13 were not experimentally validated in Ref.~\cite{welz2015new}, contributing to significant uncertainty in the branch ratios. To improve prediction accuracy, this model adjusted the branching ratios of Reactions R10-R13 based on DBD experimental results. This adjustment notably enhanced the prediction accuracy for $\rm i\text{-}C_3H_7OH$, although a slight increase in $\rm n\text{-}C_3H_7OH$ mole fraction compared to the original model was observed. This increase can be attributed to the enhanced reaction rate of $\rm n\text{-}C_3H_7O$ with $\rm HO_2$, reflecting the increased mole fraction of $\rm HO_2$ resulting from the deletion of Reaction R6.

Additionally, a formation pathway for $\rm C_2H_5OH$ was introduced based on the NUIGMech1.1 mechanism~\cite{ramalingam2021chemical}, using rate coefficients derived from self- and cross-reactions similar to those of $\rm n\text{-}/i\text{-}C_3H_7O_2$. This inclusion brought the model's prediction for $\rm C_2H_5OH$ within the experimental measurement error range, as depicted in Fig.~\ref{Validation-exp-DBD}.

Furthermore, the original model showed under-prediction of $\rm C_2H_5OOH$ compared to experimental measurements, with almost all $\rm C_2H_5OOH$ formed via the reaction between $\rm C_2H_5O_2$ and $\rm HO_2$. The increased mole fraction of $\rm HO_2$ in the updated model led to an enhanced reaction rate between $\rm C_2H_5O_2$ and $\rm HO_2$, aligning the predicted mole fraction of $\rm C_2H_5OOH$ more closely with experimental measurements. This further supports the decision to delete Reaction R6.

\begin{figure}[htbp]
	\centering
	\includegraphics[width=1.0\linewidth]{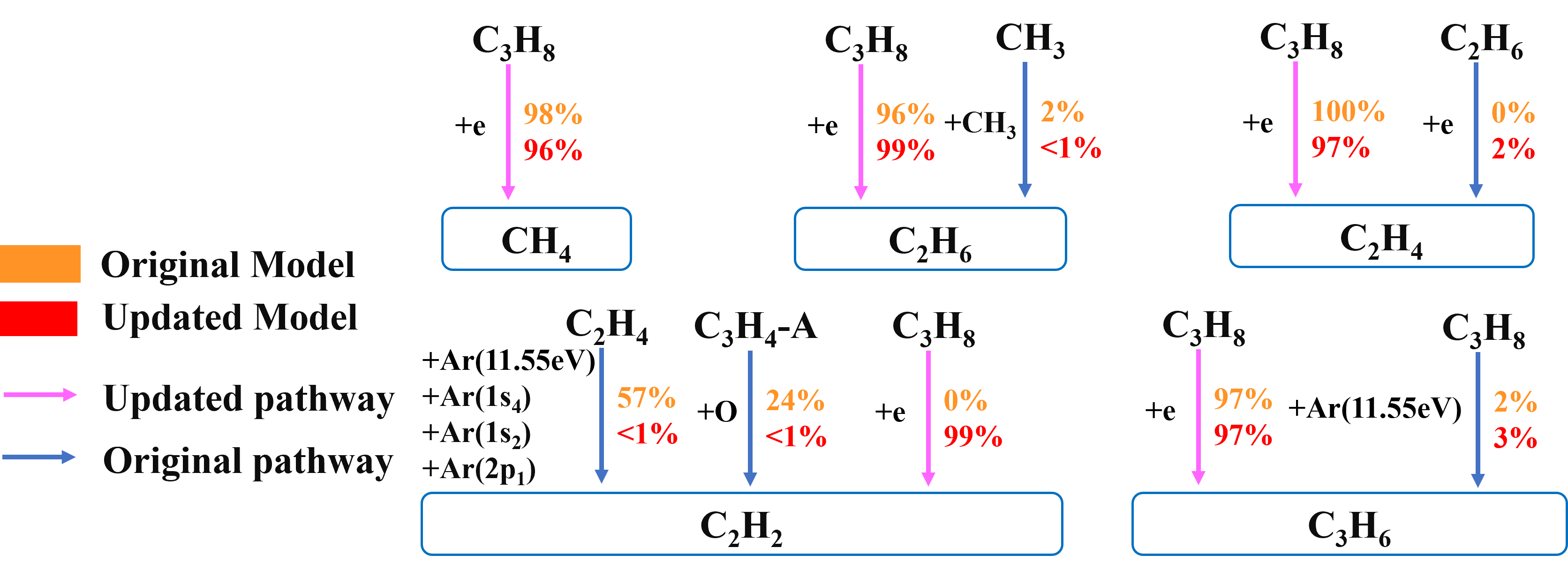}
	\caption{The pathway analysis of $\rm CH_4$, $\rm C_2H_2$, $\rm C_2H_4$, $\rm C_2H_6$ and $\rm C_3H_6$ in the updated model and the original model on the process of nCD assisted propane oxidation.}
	\label{pathway_of_CxYx}
\end{figure}

The discrepancies of $\rm CH_4$, $\rm C_2H_2$, $\rm C_2H_4$, $\rm C_2H_6$ and $\rm C_3H_6$ between experimental measurements and the original model are observed in the DBD experiment and the nCD experiment. Given the large energy deposition scales of the nCD experiment, it was chosen as the primary reference for adjusting the kinetic model. A discharge peak voltage of 5.0 kV was selected for pathway flux analysis, as shown in Fig.~\ref{pathway_of_CxYx}. In the original model, $\rm CH_4$, $\rm C_2H_4$, $\rm C_2H_6$, and $\rm C_3H_6$ are produced from the electron impact dissociation of $\rm C_3H_8$. The dissociation cross sections for $\rm C_3H_8$ were obtained from Ref.~\cite{janev2002collision}, which provided estimates based on the total experimental dissociation cross sections for both ionized and neutral fragments ($\rm e + C_3H_8 \rightarrow A + B + e$ and $\rm e + C_3H_8 \rightarrow A^+ + B + 2e$). However, the cross-section branch ratios were estimated using empirical formulas, leading to significant uncertainties. Consequently, the reaction rate coefficients for $\rm C_3H_8$ dissociation were adjusted according to experimental results.

First, in the original model, Reaction (R1) is the primary production pathway for $\rm C_3H_6$, and its prediction was overestimated across all discharge voltage scales in the nCD experiment. To improve accuracy, the cross section for Reaction (R1) was reduced by a factor of 0.8, which enhanced the prediction accuracy for $\rm C_3H_6$, as shown in Fig.~\ref{Validation_With_FIW}.

Second, reaction (R2) is responsible for the formation of $\rm C_2H_6$. The original model significantly under-predicted $\rm C_2H_6$ in the nCD experiment. An adjustment factor of 12 was applied to the cross section for Reaction (R2), aligning the $\rm C_2H_6$ predictions with the experimental results.

Furthermore, reaction (R3) is the primary pathway for the formation of both $\rm CH_4$ and $\rm C_2H_4$. The plasma sub-mechanism for $\rm CH_4$ was adopted from Ref.~\cite{mao2019numerical,mao2019methane} with experimental validation, indicating that the $\rm CH_4$ consumption rates in the original model were accurate. Despite this, both $\rm CH_4$ and $\rm C_2H_4$ were over-predicted in the DBD experiment, and $\rm CH_4$ was also over-predicted in the nCD experiment. This suggested that the cross section for Reaction (R3) was overestimated. An adjustment to the cross section for Reaction (R3) was made to achieve better agreement for $\rm CH_4$ in the nCD experiment and $\rm C_2H_4$ in the DBD experiment.

Besides, A discrepancy in the prediction of $\rm C_3H_4$ between the original model and the DBD experiment was noted. Reaction (R4) is responsible for 93\% of the formation of $\rm C_3H_4\text{-}A$ and $\rm C_3H_4\text{-}P$. Thus, the rate of Reaction (R4) was adjusted to better match the DBD experimental results.

Finally, the original model under-predicted $\rm C_2H_2$ in both the nCD and DBD experiments, as shown in Fig.~\ref{Validation_With_FIW} and Fig.~\ref{Validation-exp-DBD}. In the original model, $\rm C_2H_2$ is generated through reactions between excited argon and $\rm C_2H_4$ and between $\rm O$ and $\rm C_3H_4\text{-}A$. However, with a mole fraction of $\rm C_2H_2$ around 600 ppm in the nCD experiment compared to a $\rm C_2H_4$ mole fraction of less than 400 ppm, $\rm C_2H_4$ alone cannot account for the $\rm C_2H_2$ formation. Similarly, $\rm C_3H_4\text{-}A$ is present at too low a concentration to significantly contribute to $\rm C_2H_2$ production. Therefore, electron-impact dissociation of $\rm C_3H_8$ (Reaction R5) was considered as a primary source for $\rm C_2H_2$. Adding Reaction (R5) to the updated model significantly improved the accuracy of $\rm C_2H_2$ predictions, with the updated model's $\rm C_2H_2$ predictions closely matching experimental results from both the nCD and DBD experiments.

As stated above, The model undergoes significant updates through pathway flux analysis informed by experiments conducted under nCD and DBD conditions. By adjusting key reaction mechanisms according to experimental data, including $\rm C_3H_6$, $\rm C_2H_6$, $\rm CH_4$, $\rm C_2H_4$, $\rm C_2H_2$, and $\rm C_3H_4$, the updated model now demonstrates markedly improved predictive accuracy. This enhancement is particularly evident in its ability to better replicate experimental observations in both nCD and DBD experiments.

\subsection{The comparisons of $\rm C_3H_8$ oxidation process assisted by DBD and nCD}\label{res_comparison}

To explore the impact of the single pulse deposition energy (SPDE) on the oxidation of $\rm C_3H_8$, we have designed three distinct cases, each tailored to mimic specific energy deposition conditions. As depicted in Table~\ref{case_parameter}, these cases are denoted as Case1, Case2, and Case3, with corresponding SPDE values in the model set at $1.85\times10^{-5}$ $J/cm^3$, $9.08\times10^{-5}$ $J/cm^3$, and $3.9\times10^{-4}$ $J/cm^3$, respectively. Notably, the total deposition energy in the model is standardized at $7.36\times10^{-3}$ $J/cm^3$ by adjusting the discharge frequency and pulse number, ensuring consistency across the cases. Furthermore, the gas temperature is maintained at a constant 400 K, while the discharge current values for the DBD and nCD in the models are sourced from Ref.~\cite{chen2023kinetic} and the nCD experiment, respectively. These cases are designed to provide insights into the role of SPDE on $\rm C_3H_8$ oxidation, offering valuable comparisons under varying energy deposition conditions.

\begin{table}[t!h!]
	\caption{Initial parameters of simulation cases.}\label{case_parameter}
    \scalebox{0.85}{
	\begin{tabular}{cccc}
	\hline
	\multicolumn{1}{l}{}                                             & \multicolumn{1}{c}{Case1} & \multicolumn{1}{c}{Case2} & \multicolumn{1}{c}{Case3} \\ \hline
	Discharge type                                                   & DBD                       & nCD                       & nCD                       \\
	Peak voltage, kV                                                 & 9.0                       & 4.5                       & 5.0                       \\ \cline{2-4} 
	Discharge time, ms                                                   & \multicolumn{3}{c}{20}                                                            \\ \cline{2-4} 
	Frequency, Hz                                                    & 20000                     & 4050                      & 950                       \\
	Pulse number                                                     & 400                       & 81                        & 19                        \\
	Single pulse deposition energy in model, $J/cm^3$ & $1.85\times10^{-5}$                   & $9.08\times10^{-5}$                   & $3.9\times10^{-4}$                    \\ \cline{2-4} 
	Total deposition energy in model, $J/cm^3$        & \multicolumn{3}{c}{$7.36\times10^{-3}$}                                                       \\
	Gas temperature, K                                               & \multicolumn{3}{c}{400}                                                           \\
	Pressure, Pa                                                     & \multicolumn{3}{c}{4000}                                                          \\
	Discharge mixture                                                & \multicolumn{3}{c}{$\rm C_3H_8/O_2/Ar$ 4:20:76}                                            \\ \hline
	\end{tabular}
	}
\end{table}

\begin{figure}[t!h!]
	\centering
	\begin{subfigure}{0.47\linewidth}
		\centering
		\includegraphics[width=1.0\textwidth]{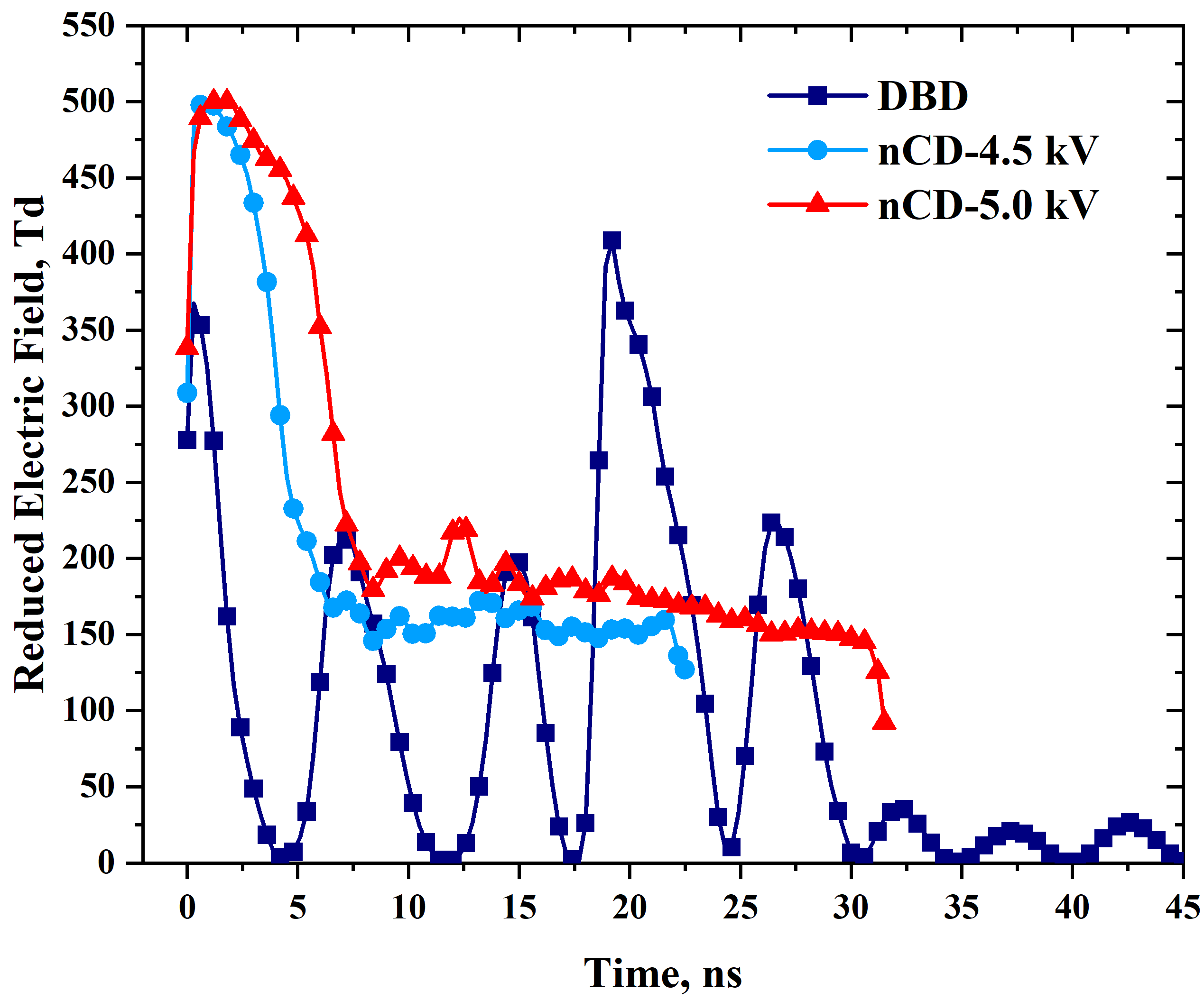}
    \end{subfigure}
	\begin{subfigure}{0.47\linewidth}
		\centering
		\includegraphics[width=1.0\textwidth]{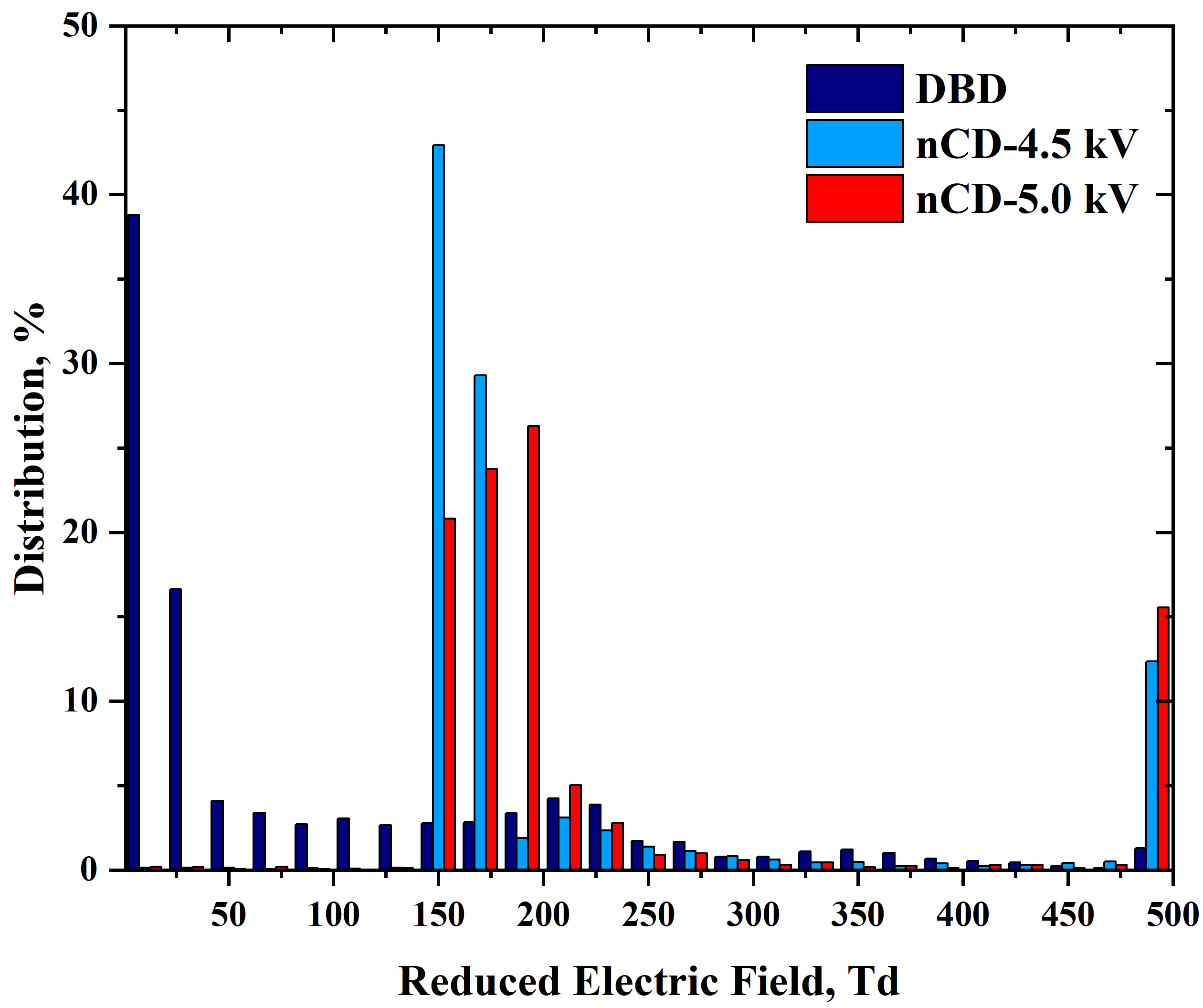}
	\end{subfigure}
	\caption{The comparison of average reduced electric field and reduced electric field distribution in DBD and nCD.}
	\label{reduced_field}
\end{figure}

In Fig.~\ref{reduced_field}(a), the average reduced electric field evolution for DBD and nCD is depicted, where the reduced electric field is averaged over all discharge pulses. Notably, the average reduced electric field for nCD experiences a rapid increase to 450-500 Td before 3 ns, followed by a dramatic decrease to 150-200 Td at 8 ns, maintaining this level until the end of the discharge. In contrast, the average reduced electric field for DBD fluctuates repeatedly from 0 Td to 400 Td. As illustrated in Fig.~\ref{reduced_field}(b), the reduced electric field distribution for nCD is concentrated at 150-200 Td and 450-500 Td. In comparison, the reduced electric field distribution for DBD predominantly ranges from 0-25 Td and 50-250 Td, accounting for approximately 55\% and 35\%, respectively, indicating a lower reduced electric field distribution compared to nCD.

\begin{figure}[htbp]
	\centering
	\includegraphics[width=0.5\linewidth]{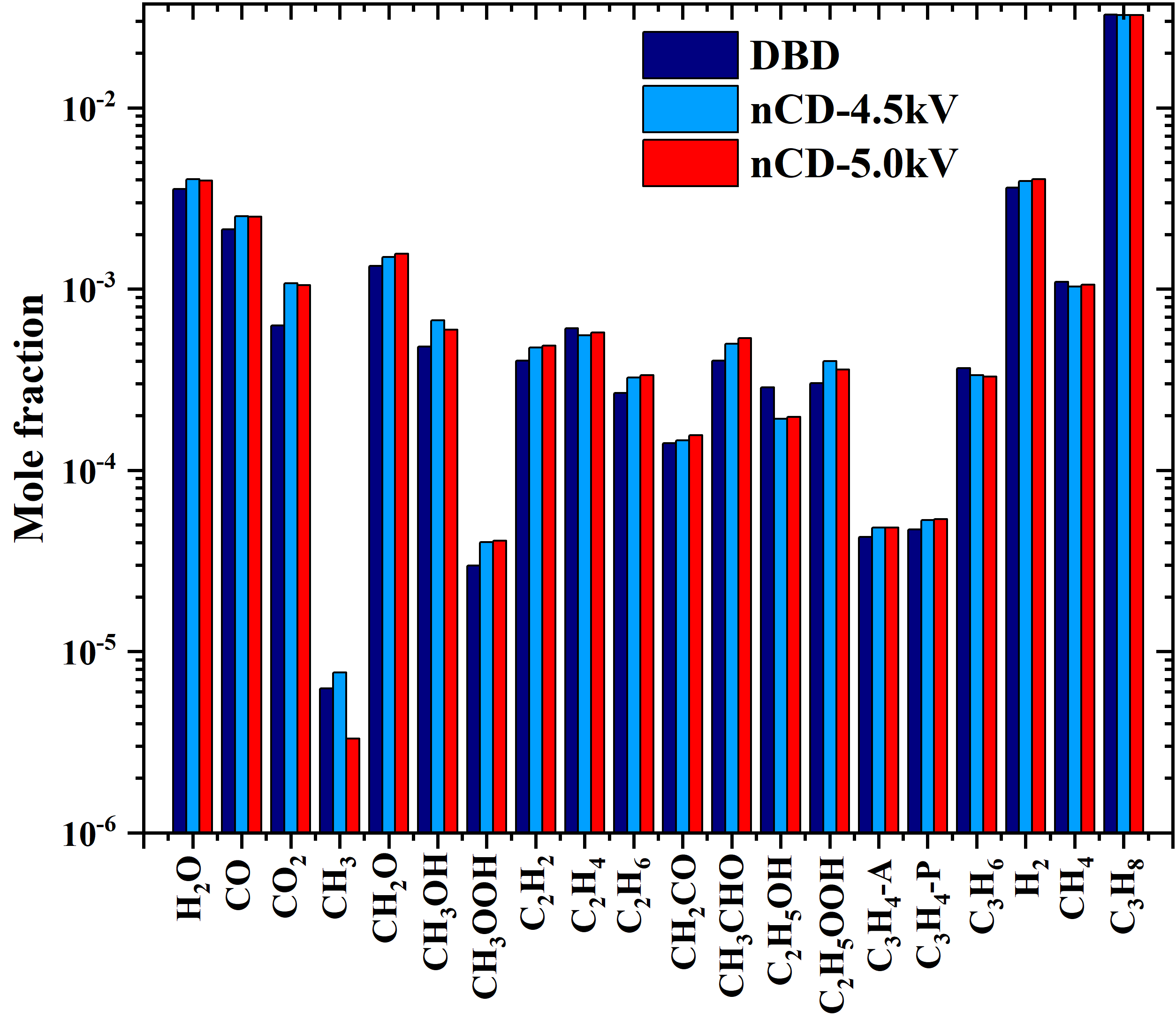}
	\caption{Comparison of species denisty in $\rm C_3H_8$ oxidation process assisted by DBD and nCD.}
	\label{comparison_of_density}
\end{figure}

\begin{figure}[t!h!]
	\centering
	\begin{subfigure}{0.47\linewidth}
		\centering
		\includegraphics[width=1.0\textwidth]{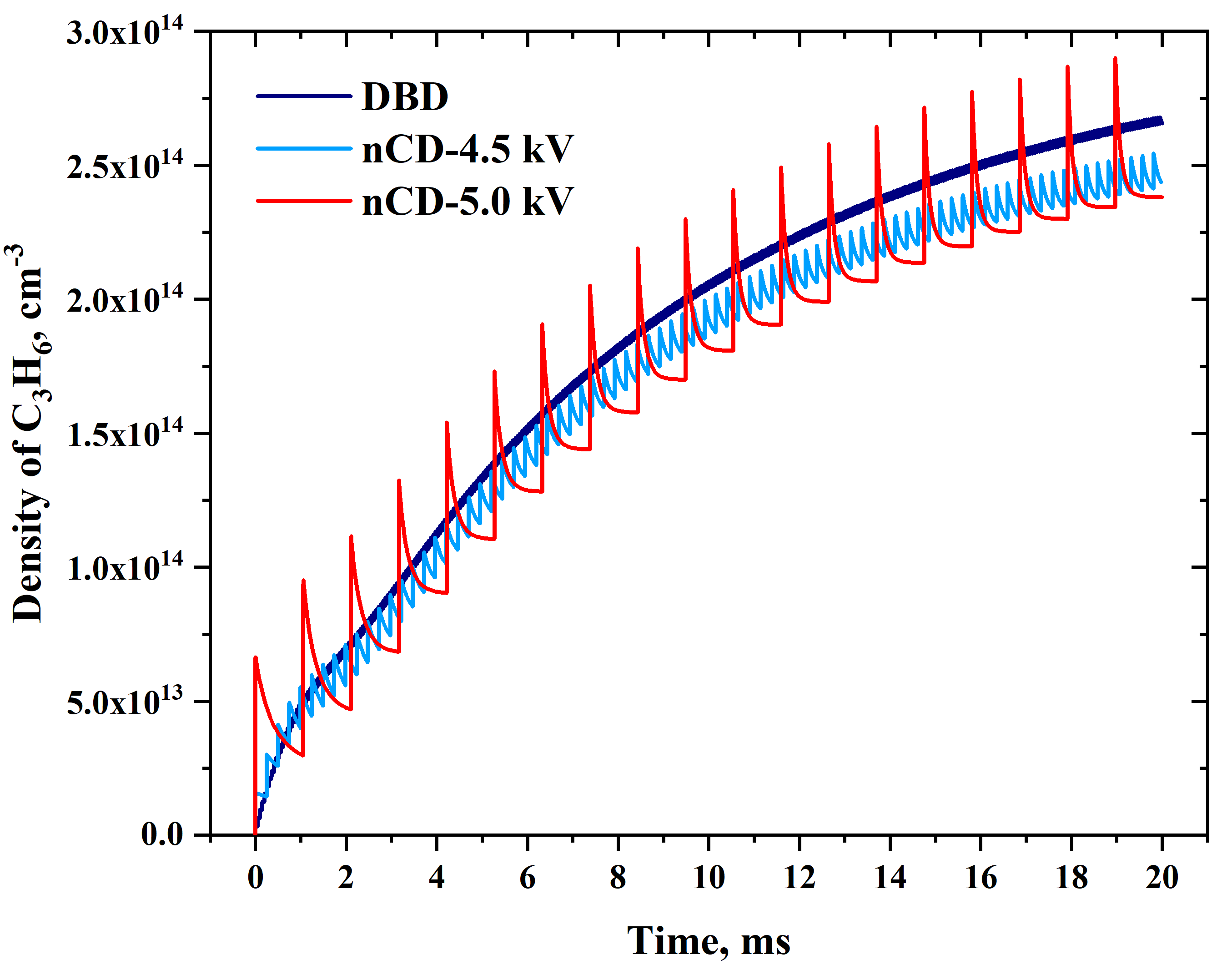}
    \end{subfigure}
	\begin{subfigure}{0.47\linewidth}
		\centering
		\includegraphics[width=1.0\textwidth]{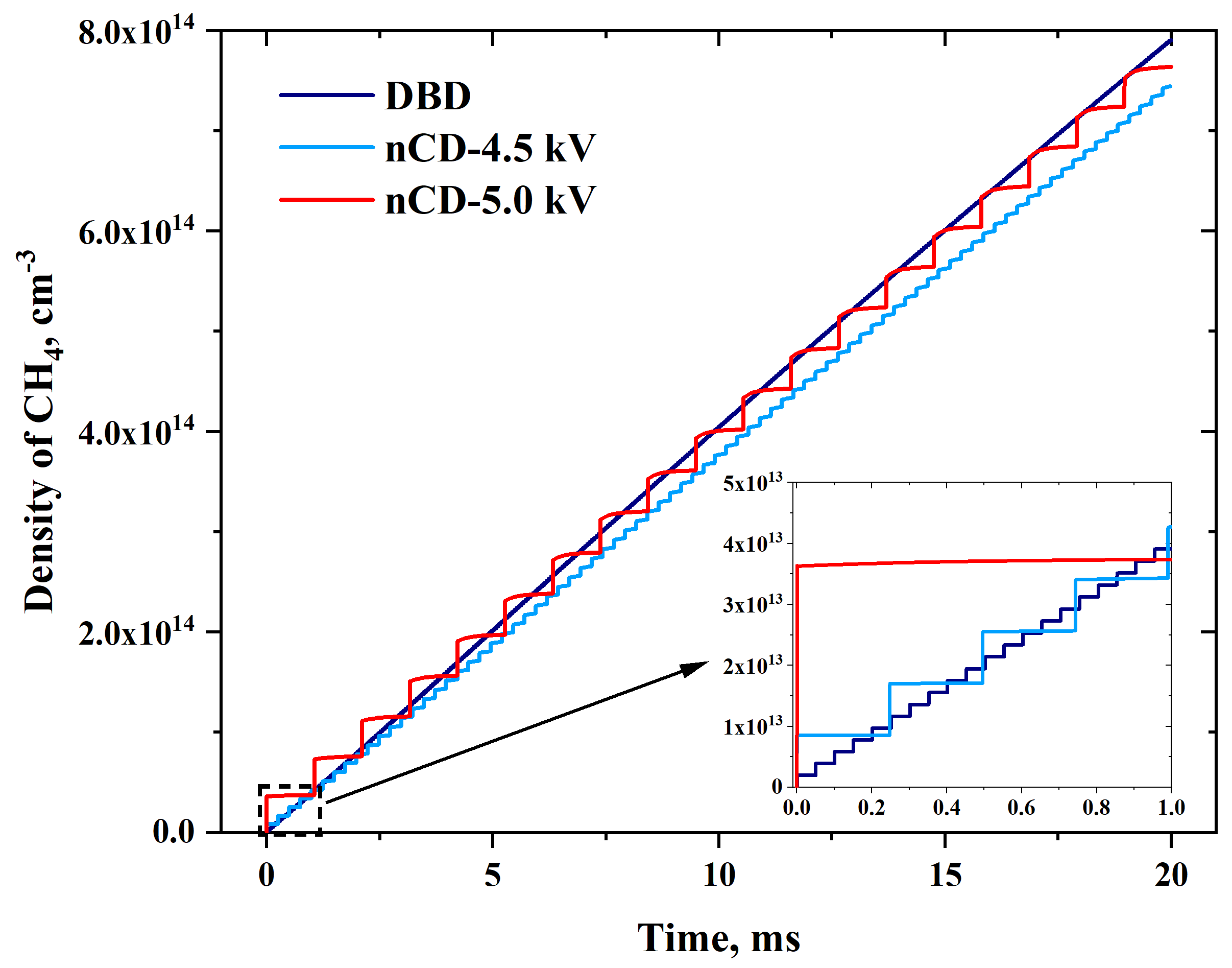}
	\end{subfigure}
	\caption{$\rm C_3H_6$ and $\rm CH_4$ time evolutions in $\rm C_3H_8$ oxidation process assisted by DBD and nCD.}
	\label{density_evolutions}
\end{figure}

Fig.~\ref{comparison_of_density} presents a comparison of species density in the $\rm C_3H_8$ oxidation process assisted by DBD and nCD. Interestingly, the density of species such as $\rm H_2O$, CO, $\rm CO_2$, $\rm C_2H_2$, and $\rm C_2H_4$ in the $\rm C_3H_8$ oxidation process remains similar between DBD and nCD, despite varying SPDE values. This suggests that SPDE has minimal influence on the $\rm C_3H_8$ oxidation process when the total deposition energy remains constant. Furthermore, Fig.~\ref{density_evolutions} highlights the time evolution of $\rm C_3H_6$ and $\rm CH_4$ in the $\rm C_3H_8$ oxidation process assisted by DBD and nCD. Both $\rm C_3H_6$ and $\rm CH_4$ are produced during nanosecond pulse discharge and subsequently consumed in the discharge gaps. Notably, the increase in $\rm C_3H_6$ and $\rm CH_4$ during a given period correlates with the rise in SPDE. These observations collectively suggest that the $\rm C_3H_8$ oxidation process is primarily controlled by total deposition energy and displays limited correlation with the specific discharge type, be it DBD or nCD.


\begin{figure}[htbp]
	\centering
	\includegraphics[width=1.0\linewidth]{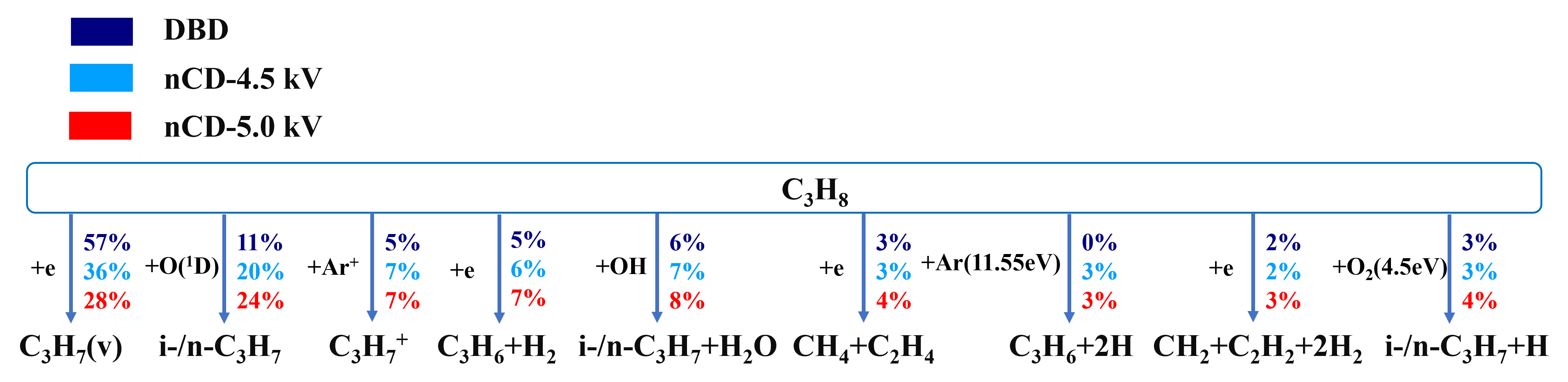}
	\caption{$\rm C_3H_8$ consumption pathways analysis in DBD and nCD.}
	\label{C3H8_nCD_DBD}
\end{figure}

As depicted in Fig.~\ref{C3H8_nCD_DBD}, the consumption pathways of $\rm C_3H_8$ in DBD and nCD environments were analyzed. In DBD, 57\% of $\rm C_3H_8$ is consumed through vibrationally excited reactions, attributed to the concentrated reduced electric field distribution of 0-25 Td, which enhances efficient vibrational excitation. Conversely, in nCD, a lower proportion of $\rm C_3H_8$ is consumed via vibrationally excited reactions, with a higher contribution from species such as $\rm O(^1D)$, $\rm Ar^+$, $\rm e$, and OH. $\rm C_3H_8(v)$ is quenched back to $\rm C_3H_8$ through relaxation reactions, contributing thermally to the process. Although a significant portion of $\rm C_3H_8$ in DBD is excited to $\rm C_3H_8(v)$, the energy deposited into $\rm C_3H_8(v)$ remains relatively low due to the limited vibrational excitation energy. In DBD, the majority of energy deposition occurs within the 100-250 Td range, similar to processes occurring at 150-200 Td and 500 Td in nCD. Consequently, the species density in the $\rm C_3H_8$ oxidation process, whether assisted by DBD or nCD, exhibits comparable values. Additionally, the time evolutions of $\rm C_3H_6$ and $\rm CH_4$ densities in both nCD and DBD environments display similar tendencies.

\section{Conclusions}\label{conclusions}

This study investigates the kinetics of plasma-assisted propane oxidation in nanosecond pulse discharge. A zero-dimensional plasma-combustion coupled solver equipped with sensitivity and pathway flux analysis modules was developed, based on the ZDPlaskin package and a custom-built combustion chemistry code. A kinetic model was proposed and validated against the nCD experiment using gas chromatography and the nanosecond DBD experiment using synchrotron photoionization mass spectrometry. The updated model demonstrates predictions for 15 species falling within the range of DBD experimental measurement errors, with significantly improved prediction accuracy for 17 species. Notably, the updated model exhibits good prediction accuracy with the nCD experiment for $\rm C_3H_8$, $\rm H_2$, $\rm CH_4$, and $\rm C_2H_6$ under large deposition energy scales. However, the model-predicted fractions of $\rm CO_2$, $\rm C_2H_4$, and CO are 55\%, 45\%, and 45\% lower than those measured in the nCD experiment, respectively.

Several modifications to the reactions were made based on experiments and pathway flux analysis. This included fitting the reaction rate coefficients of $\rm i \text{-}/n\text{-}C_3H_7O_2$ with $\rm HO_2$ (R7, R8) due to the large uncertainty of the analogy method. Additionally, the deletion of the erroneous addition reaction $\rm CH_3OH+O_2\leftrightarrow CH_3O+HO_2$ resulted in improved model predictions for $\rm CH_3OH$. The supplementation of the $\rm CH_3O$ sub-mechanism from the HP-Mech v3.3 mechanism enhanced the updated model's prediction accuracy for $\rm CH_2O$. Furthermore, adjustments to the $\rm C_3H_8$ dissociation reaction rate coefficients, owing to large uncertainties, greatly improved the prediction accuracy for $\rm H_2$, $\rm CH_4$, $\rm C_2H_6$, $\rm C_2H_2$, and $\rm C_3H_6$.

Comparisons of $\rm C_3H_8$ oxidation processes assisted by DBD and nCD under varying SPDE but with the same total deposition energy revealed that the process is primarily controlled by total deposition energy, demonstrating minimal influence from SPDE and displaying limited correlation with the specific discharge types, DBD and nCD.

Overall, these findings illustrated the intricate interplay between plasma and combustion chemistry, offering valuable insights into the mechanisms of plasma-assisted propane oxidation.


\section*{Acknowledgments}
The work is supported by the National Natural Science Foundation of China (No.52277168, No.52025064, No.52207151), the National Key Research and Development Program of China (No.2023YFB4005700, No.2023YFB4005705, and No.2023YFB4005702-03), the University Synergy Innovation Program of Anhui Province (No.GXXT-2022025), and the independent project of the Energy Research Institute of Hefei Comprehensive National Science Center (Anhui Energy Laboratory) (No.22KZZ525, No.23KZS402, No.22KZS301, and No.22KZS304). The authors are thankful to the young research group in Atelier des Plasmas for fruitful discussions.

\bibliographystyle{elsarticle-num-names}
\bibliography{GA_Li}


%
%
%
\end{document}